\documentclass{article}
\usepackage{amsmath}
\usepackage{amssymb}
\usepackage{amsfonts}
\usepackage{latexsym,longtable}
\usepackage[colorlinks,linkcolor=blue,anchorcolor=blue,citecolor=green,CJKbookmarks=True]{hyperref}

\begin{document}
\newcommand{\B}{{\cal B}}
\newcommand{\D}{{\cal D}}
\newcommand{\E}{{\cal E}}
\newcommand{\C}{{\cal C}}
\newcommand{\F}{{\cal F}}
\newcommand{\A}{{\cal A}}
\newcommand{\Hh}{{\cal H}}
\newcommand{\Pp}{{\cal P}}
\newcommand{\G}{{\cal G}}
\newcommand{\Z}{{\bf Z}}
\newcommand{\T}{{\cal T}}
\newcommand{\ZZ}{{\mathbb{Z}}}
\newcommand{\qed}{\hphantom{.}\hfill $\Box$\medbreak}
\newcommand{\proof}{\noindent{\bf Proof \ }}
\renewcommand{\theequation}{\thesection.\arabic{equation}}
\newtheorem{theorem}{Theorem}[section]
\newtheorem{lemma}[theorem]{Lemma}
\newtheorem{corollary}[theorem]{Corollary}
\newtheorem{remark}[theorem]{Remark}
\newtheorem{example}[theorem]{Example}
\newtheorem{definition}[theorem]{Definition}
\newtheorem{construction}[theorem]{Construction}

%%%%%%%%%%%%%%%%%%%%%%%%%%%%%%%%%%%%%%%%%%%%%%%%%%%%%%%%%%%%%%%%%%%%%%%%

\medskip
\title{Combinatorial Constructions of Optimal
 $(m, n,4,2)$ Optical Orthogonal Signature Pattern
Codes\thanks{Research supported by NSFC grants 11222113, 11431003, and a project funded by the priority academic program development
of Jiangsu higher education institutions.}}

 \author{{\small  Jingyuan Chen, Lijun Ji\thanks{Corresponding author}\ and Yun Li} \\
 {\small Department of Mathematics, Soochow University, Suzhou
 215006, China}\\
 {\small E-mail: jilijun@suda.edu.cn}\\
 }

\date{}
\maketitle
\begin{abstract}
\noindent \\ Optical orthogonal signature pattern codes
(OOSPCs) play an important role in a novel type of optical code division multiple access (OCDMA)
network for 2-dimensional image transmission. There is a one-to-one correspondence
between an $(m, n, w, \lambda)$-OOSPC and a $(\lambda+1)$-$(mn,w,1)$ packing design admitting a point-regular automorphism group isomorphic to $\mathbb{Z}_m\times \mathbb{Z}_n$.
In 2010, Sawa gave the first infinite class of $(m, n, 4, 2)$-OOSPCs by using $S$-cyclic Steiner quadruple systems.
In this paper, we use various combinatorial designs such as strictly $\mathbb{Z}_m\times \mathbb{Z}_n$-invariant $s$-fan designs, strictly $\mathbb{Z}_m\times \mathbb{Z}_n$-invariant $G$-designs and rotational Steiner quadruple systems to present some constructions for $(m, n, 4, 2)$-OOSPCs. As a consequence, our new constructions yield more infinite families of optimal $(m, n, 4, 2)$-OOSPCs. Especially, we see that in some cases an
optimal $(m, n, 4, 2)$-OOSPC can not achieve the Johnson bound. We also use Witt's inversive planes to obtain optimal  $(p, p, p+1, 2)$-OOSPCs for all primes $p\geq 3$.

\medskip

\noindent {\bf Keywords}: Automorphism group,  packing design, optical orthogonal code,  optical orthogonal signature pattern
code, spatial optical CDMA.

\end{abstract}

%%%%%%%%%%%%%%%%%%%%%%%%%%%%%%%%%%%%%%%%%%%%%%%%%%%%%%%%%%%%%%%%%%%%%%%%%

\section{Introduction}

Optical code division multiple access (OCDMA) allows many nodes, which technically collide with one another,
to transmit and be accessed simultaneously and dynamically
with no waiting time at the same wavelength \cite{DS1983}, \cite{TNO1985}, \cite{PSF1986},
\cite{Salehi1993}.
Spatial OCDMA, an extension of OCDMA to a two-dimensional (2-D) space coding for
image transmission and multiple access, can exploit the inherent
parallelism of optics. The beauty of parallelism is that a light beam can carry the
information of a 2-D array of pixels of a binary-digitized image, and hence the 2-D array of pixels of an image can be transmitted and processed simultaneously through multicore fiber without parallel-to-serial conversion. For details on fundamentals and applications of OCDMA and spatial OCDMA, the reader is referred to books \cite{Kitayama2014}, \cite{Prucnal2006}, \cite{YR2007}.
The approach, which combines the features of optical image
processing with multiple access inherent in OCDMA, has been proposed and
demonstrated for encoding two dimensional pixel arrays by Kitayama and colleagues \cite{Kitayama1994,KNIK1997}. This 2-D image multiplexing has
many applications such as transmission of medical images, parallel optical
interconnections between processors and memory in high performance
computing, etc \cite{Prucnal2006}, \cite{YR2007}.  And the spatial OCDMA provides higher
throughput comparing with the
traditional OCDMA \cite{Kitayama2014}.

In spatial OCDMA,  each magnified 2-D bit plane, comprising individual pixels of an image,
is encoded using a $(0,1)$ 2-D
matrix called a 2-D optical orthogonal signature pattern (OOSP). The encoded image is constructed by taking
the Hadamard product of the overlapping matrix elements of the magnified bit plane
and OOSP. Each node on the network uses a unique OOSP to encode the planar data
from its input image. When a receiver which possesses the signature pattern receives the encoded signal,
it extracts the input image from the received encoded signal by correlating the received encoded signal and its own OOSP with
a specific signature function. For more details, the interested readers may refer to \cite{HHR1995}, \cite{Kitayama1994}.

As pointed out in \cite{Kitayama1994}, one of the keys to spatial OCDMA is the methodology
in the construction of the 2-D OOSPs.
The construction of the OOSPs for 2-D data encoding follows many of the same principles typical of optical CDMA codes
including code orthogonality and large cardinality. In addition, each OOSP
is distinguishable from space-shifted versions of themselves (auto-correlation) on a 2-D plane and any
two different OOSPs in a set are distinguishable from each other (cross-correlation), even
with the existence of vertical or horizontal space shifts in the plane \cite{HHR1995}, \cite{Kitayama1994}.
The constraints require that the correlation is much lower than the weight (the number of ``1") of the OOSP.
Now we introduce the formal definition of an
optical orthogonal signature pattern code.

Let $m,n,w,\lambda$ be positive integers with $mn>w\geq \lambda$. An optical orthogonal signature pattern code with $m$ wavelengths, time-spreading length $n$, constant weight $w$ and the maximum collision parameter $\lambda$, or briefly $(m, n, w, \lambda)$-OOSPC, is a family, ${\mathcal C}$,
of $m\times n$ (0, 1)-matrices (codewords) with constant Hamming weight $w$ (i.e., the number of ones) such that the
following correlation properties hold:

(1) (Auto-Correlation Property)
$$\sum\limits_{i=0}^{m-1}\sum\limits_{j=0}^{n-1}a_{i,j}a_{i\oplus_m \delta,j\oplus_n\tau}\leq\lambda$$
for any matrix $A = (a_{i, j} )\in {\cal C}$ $(0 \leq i \leq m-1$, $0\leq j \leq n- 1)$ and any integers $\delta$ and $\tau$ with $0\leq \delta<m$, $0\leq \tau <n$ and $(\delta,\tau)\neq (0,0)$;

(2) (Cross-Correlation Property)
$$\sum\limits_{i=0}^{m-1}\sum\limits_{j=0}^{n-1}a_{i,j}b_{i\oplus_m \delta,j\oplus_n\tau}\leq\lambda$$
for any two distinct matrices $A = (a_{i, j}), B = (b_{i, j} ) \in  {\cal C}$ $(0 \leq i \leq m-1$, $0\leq j \leq n- 1)$,
and any integers $\delta$ and $\tau$ with $0\leq \delta<m$ and $0\leq \tau <n$, where $\oplus_m$ (resp. $\oplus_n$) denotes addition modulo $m$ (resp. modulo $n$) and equality holds in each of the two inequalities for
at least one instance.

When the auto-correlation property is replaced by $\sum_{i=0}^{m-1}\sum_{j=0}^{n-1}a_{i,j}a_{i,j\oplus_n\tau}\leq\lambda$ ($0<\tau<n$) and the cross-correlation property is replaced by $\sum_{i=0}^{m-1}\sum_{j=0}^{n-1}a_{i,j}b_{i,j\oplus_n\tau}\leq\lambda$ ($0\leq \tau<n$), this defines a two-dimensional optical orthogonal code (2-D $(m \times n, w, \lambda)$-OOC). Clearly, an OOSPC is a special 2-D OOC, for example, see \cite{YK1997} for other variations of 2-D OOCs. Many constructions for 2-D $(m \times n, w, \lambda)$-OOCs have been given, see \cite{AM2009}, \cite{BJ-JCD-appear},  \cite{CWS2009}, \cite{FC2011}, \cite{HC2012}, \cite{OGKEB2012}, \cite{SYWX2006}, \cite{YK1997}.
Note that in the particular case where $m = 1$
and $n = v$, a 2-D $(m, n, w, \lambda)$-OOSPC  is nothing else than a one-dimensional $(v, w, \lambda)$
optical orthogonal code (briefly, 1-D $(v, w, \lambda)$-OOC). For details on 1-D OOCs, the reader is referred to  \cite{CSW1989}, \cite{CDS1999}, \cite{ML1998}, \cite{SB1989}.
So far,
many constructions of 1-D OOCs
with maximum size and many results have been made,
for example, \cite{AB2004}, \cite{BW1987}, \cite{Buratti2002}, \cite{BP2010},  \cite{CJ2004}, \cite{CZ1998}, \cite{FMY2001}, \cite{SB1989}, \cite{YYL2011}.

Throughout this paper we always denote by $\mathbb{Z}_n$ the additive
group of integers modulo $n$.
For each $(0, 1)$-matrix $A = (a_{i j} ) \in  \C$, whose rows are
indexed by $\mathbb{Z}_m$ and columns are indexed by $\mathbb{Z}_n$, we define
$X_A = \{(i, j) \in \mathbb{Z}_m \times \mathbb{Z}_n : a_{ij} = 1\}$. Then, $\F = \{X_A : A \in \C\}$
is a set-theoretic representation of an $(m, n, w, \lambda)$-OOSPC.
Thus, an $(m,n,w,\lambda)$-OOSPC is a set $\F$ of
$w$-subsets of $\mathbb{Z}_m \times \mathbb{Z}_n$ in which each $w$-subset $X$ corresponds to a signature pattern
 $(a_{i j})$ such that $a_{i j} = 1$ if and only if
$(i, j )\in X$, where the two correlation properties are given as follows:

$(1^{'})$ (Auto-Correlation Property)
$$|X \cap (X + (\delta, \tau))| \leq \lambda$$
for each $X \in \F$ and every
$(\delta, \tau) \in  \mathbb{Z}_m \times \mathbb{Z}_n\setminus \{(0, 0)\}$;

$(2^{'})$ (Cross-Correlation Property)
$$|X \cap (Y + (\delta, \tau))| \leq \lambda$$
for any distinct
$X, Y \in \F$ and every $(\delta, \tau) \in  \mathbb{Z}_m \times \mathbb{Z}_n$.

 The number of codewords in an OOSPC is called the {\em size} of
the OOSPC. For given integers $m, n, w$ and $\lambda$, let $\Theta(m, n, w, \lambda)$
be the largest possible size among all $(m, n, w, \lambda)$-OOSPCs.
An $(m, n, w, \lambda)$-OOSPC with size $\Theta(m, n, w, \lambda)$ is said to be
{\em optimal}. Based on the Johnson bound \cite{Johnson1962} for constant weight codes,
an upper bound  on the largest possible size $\Theta(m, n, w, \lambda)$ of an $(m, n, w, \lambda)$-OOSPC was given below:
\begin{equation}
\label{2D-JohnsonBound}
\Theta(m, n, w, \lambda)\leq J(m, n, w, \lambda)=\left \lfloor\frac{1}{w} \left \lfloor\frac{mn-1}{w-1}
\left \lfloor\frac{mn-2}{w-2}\left \lfloor\cdots \left \lfloor\frac{mn-\lambda}{w-\lambda}
\right \rfloor\cdots \right \rfloor\right \rfloor\right \rfloor\right \rfloor.
\end{equation}

%As stated in Section II, an $(m, n, w, \lambda)$-OOSPC with $b$ codewords produces a $(\lambda+1)$-$(mn,w,1)$ packing with $bmn$ blocks,
%which is equivalent to a binary code of length $mn$,
%size $bmn$, constant weight $w$, and minimum Hamming distance at least $2(w-\lambda)$ (denoted by an $(mn,w,\geq 2w-2\lambda)$ constant-weight binary code).
%Much work has been done on lower bound on the largest possible size $A(u,w,d)$ of a $(u,w,d)$ constant-weight binary code. It is well known that $A(u,w,d)$ is upper bounded by Johnson bound \cite{Johnson1962}. However, not much $A(u,w,d)$ have been determined. To our knowledge,  the sizes $A(u,w,2w-2)$ have been determined for some small $w$; for any prime power $q$ and some special $u$, $A(u,q,2q-2)$ have been determined by using finite projective planes and affine planes \cite{CD2007}. Hence, an $(m, n, w, \lambda)$-OOSPC with $J(m, n, w, \lambda)$ codewords yields an optimal constant-weight binary code or an nearly optimal constant-weight binary code. From the known results on optimal $(u,w,2w-2\lambda)$ constant-weight binary codes, it is hard to consider general $w$ for OOSPCs. In this paper, we are mainly concerned with $w=4$.

When $m$ and $n$ are coprime, it has been shown in \cite{YK1996} that
an $(m, n, w, \lambda)$-OOSPC is actually a 1-D $(mn, w, \lambda)$-OOC.
However, when $m$ and $n$ are not coprime, the problem of constructing optimal $(m, n, w, \lambda)$-OOSPCs becomes difficult.
Some infinite classes of optimal $(m, n, w, 1)$-OOSPCs have been given for specific values of $m,n,w$, see \cite{Buratti2001}, \cite{PC2014}, \cite{PC2015}, \cite{SK2009}, \cite{YK1996}. To our knowledge, the only known optimal OOSPCs with $\lambda\geq 2$ were obtained by Sawa \cite{Sawa2010}. He showed that there is an optimal $(2^{\epsilon}x,n,4,2)$-OOSPC where
$\epsilon\in \{1, 2\}$,
and each prime factor of $x, n$ is less than 500000 and
congruent to 53 or 77 modulo 120 or belongs to $S =
\{5, 13,17, 25, 29, 37, 41, 53$, $61, 85, 89, 97, 101, 113, 137,
149, 157, 169, 173, 193, 197, 229, 233, 289, 293, 317\}$. In this paper,
We use various combinatorial structures to present more infinite families of optimal $(m,n,4,2)$-OOSPCs.

This paper is organized as follows. In Section II,  a correspondence between an $(m,n,w,\lambda)$-OOSPC and a strictly $\mathbb{Z}_m\times \mathbb{Z}_n$-invariant $(\lambda+1)$-$(mn,w,1)$ packing design is described. Based on this correspondence we give an improved upper bound on $\Theta(m, n, 4, 2)$ by analyzing the leave of a strictly $\mathbb{Z}_m\times \mathbb{Z}_n$-invariant $3$-$(mn,4,1)$ packing design. We also construct an optimal $(p,p,p+1,2)$-OOSPC from an inversive plane of prime order $p$. Section III introduces a concept of strictly $\mathbb{Z}_m\times \mathbb{Z}_n$-invariant $G(\frac{m}{e},en,4,3)$ design, from which we can obtain a strictly $\mathbb{Z}_m\times \mathbb{Z}_n$-invariant 3-$(mn,4,1)$ packing design. We also use a cyclic SQS$(m)$ to construct a strictly $\mathbb{Z}_m\times \mathbb{Z}_n$-invariant $G(m,n,4,3)$ design. In Section IV, we give a recursive construction for
strictly $\mathbb{Z}_m\times \mathbb{Z}_n$-invariant $G^*(\frac{m}{e},en,4,3)$ design. Section V uses $1$-fan designs to present a recursive construction for strictly $\mathbb{Z}_m\times \mathbb{Z}_n$-invariant $G(m,n,w,3)$ design. Based on known $S$-cyclic SQSs and rotational SQSs, many new optimal $(m,n,4,2)$-OOSPCs are established in Section VI. Finally, Section VII gives a brief conclusion. Our main results are summarized in Table I.

\section{Combinatorial characterization}

In this section, we describe a correspondence between an $(m,n,k,\lambda)$-OOSPC and a strictly $\mathbb{Z}_m\times \mathbb{Z}_n$-invariant $(\lambda+1)$-$(mn,k,1)$ packing design. Based on this correspondence we give an improved upper bound on $\Theta(m, n, 4, 2)$ and construct an optimal $(p,p,p+1,2)$-OOSPC for any prime $p$.

Let $t,w, n$  be positive integers. A $t$-$(n,w,1)$ {\it packing design} consists of an $n$-element set
$X$ and a collection $\B$ of $w$-element subsets of $X$, called {\it
blocks}, such that every $t$-element subset of $X$ is contained in
at most one block. A $3$-$(n,4,1)$ packing design is called a {\it packing
quadruple system} and denoted by PQS$(n)$.
When ``at most" is replaced by ``exactly", this defines a {\em Steiner system}, denoted by $S(t,w,n)$. An $S(2,3,n)$ is called a Steiner triple system and denoted by STS$(n)$. An $S(3,4,n)$ is called a Steiner quadruple system and denoted by SQS$(n)$. It is well known that there is an SQS$(n)$ if and only if $n\equiv 2,4\pmod 6$ \cite{Hanani1960}.

A $t$-$(n,w,1)$ packing design is optimal if it has the largest possible number $D(n,w,t)$
of blocks. It is well known \cite{Johnson1962} that $$D(n,w,t)\leq \left\lfloor\frac{n}{w}\left \lfloor\frac{n-1}{w-1} \left\lfloor\cdots \left\lfloor \frac{n-t+1}{w-t+1}\right\rfloor\cdots \right\rfloor\right\rfloor\right\rfloor.$$
For $t=2$ and $w\in \{3,4\}$, the numbers $D(n,w,2)$ have been completely determined,  there is also much work on $D(n,5,2)$, see \cite{CD2007}.
For $t\geq 3$, only numbers $D(n,4,3)$ have been completely determined \cite{BJ-DCC-appear}.

An {\em automorphism} $\sigma$ of a packing design $(X,{\cal B})$ is
a permutation on $X$ leaving ${\cal B}$ invariant.
All  automorphisms of a packing design form a group, called the {\em full automorphism group} of the packing design.
Any subgroup of the full automorphism group is called an {\em automorphism group} of the packing design. Let $G$ be
an automorphism group of a packing design. For any block $B$ of the packing design, the subgroup $$\{\sigma\in G: B^{\sigma}=B\}$$
is called the {\em stabilizer} of $B$ in $G$, where $B^{\sigma}$ stands for $\sigma$ acting on $B$. The orbit of $B$
under $G$ is the collection Orb$_G(B)$ of all distinct images of
$B$ under $G$, i.e.,
$$Orb_G(B) = \{B^{\sigma} : \sigma \in G\}.$$
It is clear that $\B$ can be partitioned into some orbits under $G$. An arbitrary set of representatives for each orbit of $\B$ is called
the set of base blocks of the packing design. A packing design $(X,\B)$ is said to be $G$-{\em invariant} if it admits $G$ as a point-regular automorphism group, that is,
$G$ is an automorphism group such that for any $x, y \in X$, there exists exactly one element $\sigma \in G$ such that $x^{\sigma} = y$.
In particular, a $\mathbb{Z}_n$-invariant packing design is {\em cyclic}. Moreover,
a  packing design $(X,B)$ is said to be {\em strictly $G$-invariant} if it is $G$-invariant and the stabilizer of each $B\in \B$ under $G$ equals the identity of $G$.
A strictly $G$-invariant $t$-$(n, w, 1)$ packing design is called {\em optimal} if it contains the largest possible number of base blocks.

For a $\ZZ_m\times \ZZ_n$-invariant $t$-$(mn, w, 1)$ packing design $(X,\B)$, without loss of generality we can identify $X$ with $\ZZ_m\times \ZZ_n$ and
the automorphisms can be taken as translations $\sigma_{a}$ defined by $x^{\sigma_a}=x+a$ for $x\in \ZZ_m\times \ZZ_n$, where $a\in \ZZ_m\times \ZZ_n$.
Thus, given an arbitrary family of all base blocks of a strictly $\ZZ_m\times \ZZ_n$-invariant $t$-$(mn, w, 1)$ packing design, we can obtain the packing design by successively adding $(i, j)$ to each base block, where $(i, j) \in \ZZ_m\times \ZZ_n$.

Based on the set-theoretic representation of an $(m, n, w, \lambda)$-OOSPC, the following connection is then obtained.

\begin{theorem} \emph{\cite{Sawa2010}}
\label{equivalence}
An  $(m, n, w, \lambda)$-OOSPC of size $u$ is equivalent
to a strictly $\mathbb{Z}_m\times \mathbb{Z}_n$-invariant $(\lambda + 1)$-$(mn, w, 1)$
packing design having $u$ base blocks.
\end{theorem}

%If an  $(m, n, w, \lambda)$-OOSPC is equivalent
%to a strictly $\mathbb{Z}_m\times \mathbb{Z}_n$-invariant $S(\lambda + 1, w,mn)$, then the OOSPC is said to be {\em perfect}.
Based on this connection, Sawa established a tighter upper bound on $\Theta(m,n,4,2)$ with $mn\equiv 0\pmod {24}$ than the Johnson bound.

\begin{lemma}
\emph{\cite{Sawa2010}}
\label{bound24k}
Let $m$ and $n$ be positive integers. If $mn\equiv 0\pmod {24}$ then $\Theta(m,n,4,2)\leq J(m,n,4,2)-1$.
\end{lemma}

Sawa \cite{Sawa2010} also posed an open problem: Does there exist an optimal $(6,n,4,2)$-OOSPC attaining the Johnson bound (\ref{2D-JohnsonBound}) for a positive integer $n$, not being a multiple of 4 in general? By analyzing the leave of a strictly $\mathbb{Z}_m\times \mathbb{Z}_n$-invariant PQS$(mn)$, we show that there does not exist an     $(m,n,4,2)$-OOSPC attaining the upper bound (\ref{2D-JohnsonBound}) for $m,n\equiv 0\pmod 3$ with $mn\equiv 18,36\pmod {72}$.

 The triples of $\mathbb{Z}_m\times \mathbb{Z}_n$ are partitioned into equivalence classes
called {\it orbits} of triples under the action of $\mathbb{Z}_m\times \mathbb{Z}_n$. The number of triples (resp. quadruples) contained in
an orbit is called the length of the orbit. If the length of an orbit is $mn$ then it is called {\em full}.
The set of triples not contained in any quadruple of a PQS$(v)$
is called the {\it leave} of this packing design.

 For  $(a,b)\in \mathbb{Z}_m\times \mathbb{Z}_n\setminus \{(0,0)\}$, denote $T_{(a,b)}=\{\{(0,0),(a,b),(x,y)\}:
(x,y)\in\mathbb{Z}_m\times \mathbb{Z}_n\setminus \{(0,0),(a,b)\}\}$. Clearly, $|T_{(a,b)}|=mn-2$. Each  quadruple of $\mathbb{Z}_m\times \mathbb{Z}_n$ either contains two triples in $T_{(a,b)}$ or does not contain any triple
in $T_{(a,b)}$. So, the number of triples which are from $T_{(a,b)}$ and in the leave of a strictly  $\mathbb{Z}_m\times \mathbb{Z}_n$-invariant PQS$(mn)$ has the same parity as $mn$.

\begin{lemma}
\label{Bound72k+18,36}  If $m\equiv 0\pmod 3$,  $n\equiv 0\pmod 3$ and $mn\equiv 0,18$ or $36\pmod {72}$, then $\Theta(m, n, 4, 2)\leq J(m,n,4,2)-1$.
\end{lemma}

\proof It is easy to see that the orbits generated by $\{(0,0),(0,\frac{n}{3}),(0,\frac{2n}{3})\}$,  $\{(0,0),(\frac{m}{3},\frac{n}{3}),(\frac{2m}{3},\frac{2n}{3})\}$,
 $\{(0,0),(\frac{m}{3},\frac{2n}{3}),(\frac{2m}{3},\frac{n}{3})\}$ and
  $\{(0,0),(\frac{m}{3},0),(\frac{2m}{3},0)\}$ under  $\mathbb{Z}_m\times \mathbb{Z}_n$ are short.  They have length $\frac{mn}{3}$ and they must be in the leave of a strictly $\mathbb{Z}_m\times \mathbb{Z}_n$-invariant PQS$(mn)$.
 Also, the other orbits of triples under  $\mathbb{Z}_m\times \mathbb{Z}_n$ are all full.
 Consequently, there are $(\frac{mn(mn-1)(mn-2)}{6}-\frac{4mn}{3})/mn=\frac{m^2n^2-3mn-6}{6}$ full orbits of triples.
On the other hand, since the number of triples containing
any given two points in the leave is even, there must be at least one triple
of the form $\{(0,0),(x,y),(x',y')\}$ with $(x',y')\neq (2x,2y)$ in the leave for $(x,y)\in \{(0,\frac{n}{3}),(0,\frac{2n}{3}),(\frac{m}{3},\frac{n}{3}),(\frac{2m}{3},\frac{2n}{3}),(\frac{m}{3},\frac{2n}{3}),(\frac{2m}{3},\frac{n}{3}),
(\frac{m}{3},0)$, $(\frac{2m}{3},0)\}$. Clearly, one full orbit of triples can not cover all such eight triples.
It follows that there are at most $\frac{m^2n^2-3mn-6}{6}-2$ full orbits of triples occurring in a strictly $\mathbb{Z}_m\times \mathbb{Z}_n$-invariant PQS$(mn)$. Therefore, $\Theta(m, n, 4, 2)\leq \lfloor\frac{m^2n^2-3mn-18}{24}\rfloor=J(m,n,4,2)-1$.
 \qed

We use an example to show that the improved upper bound in Lemma \ref{Bound72k+18,36} is tight. The corresponding optimal $(6,6,4,2)$-OOSPC gives an answer to the problem listed in Table I in \cite{Sawa2010}
\begin{example}
\label{Z6Z6-PQS(36)}
There exists a strictly $\mathbb{Z}_6\times \mathbb{Z}_6$-invariant PQS$(36)$ with $48$ base blocks, whose size meets the upper bound in Lemma $\ref{Bound72k+18,36}$.
\end{example}

\proof The following $48$ base blocks generate the block set of a strictly $\mathbb{Z}_6\times \mathbb{Z}_6$-invariant PQS$(36)$ over $\mathbb{Z}_6\times \mathbb{Z}_6$.
\[
\begin{array}{ll}
\multicolumn{2}{l}{\{(0,0),a,-a,(3,3)\},\ {\rm where}\ a\in\{(0,1),(1,0),(1,2),(1,4)\};}\\
\multicolumn{2}{l}{\{(0,0),b,(3,0),(0,3)+b\},\ {\rm where}\ b\in\{(0,1),(0,2),(1,0),} \\
\multicolumn{2}{l}{\hspace{4.5cm} (1,1),(1,2),(2,0),(2,1)(2,2)\};}\\
\{(0,0), (0,1), (1,0), (1,1)\}, & \{(0,0), (0,1), (1,2), (1,3)\}, \\

 \{(0,0), (0,1), (1,4), (1,5)\}, & \{(0,0), (0,1), (2,0), (2,1)\}, \\

 \{(0,0), (0,1), (2,2), (2,3)\}, & \{(0,0), (0,1), (2,4), (3,2)\}, \\

 \{(0,0), (0,1), (2,5), (4,2)\}, & \{(0,0), (0,1), (3,5), (4,3)\}, \\

 \{(0,0), (0,2), (1,0), (1,2)\}, & \{(0,0), (0,2), (1,1), (1,3)\}, \\

 \{(0,0), (0,2), (1,4), (2,1)\}, & \{(0,0), (0,2), (1,5), (2,0)\}, \\

 \{(0,0), (0,2), (2,2), (3,3)\}, & \{(0,0), (0,2), (2,3), (2,5)\}, \\

 \{(0,0), (0,2), (2,4), (4,4)\}, & \{(0,0), (0,2), (3,5), (4,0)\}, \\

 \{(0,0), (0,2), (4,1), (5,4)\}, & \{(0,0), (0,2), (4,2), (5,3)\}, \\

 \{(0,0), (1,0), (2,1), (3,1)\}, & \{(0,0), (1,0), (2,2), (5,4)\}, \\

 \{(0,0), (1,0), (2,3), (5,1)\}, & \{(0,0), (1,0), (2,4), (3,5)\}, \\

 \{(0,0), (1,0), (2,5), (5,3)\}, & \{(0,0), (1,0), (3,2), (4,2)\}, \\

 \{(0,0), (1,0), (4,1), (5,2)\}, & \{(0,0), (1,1), (2,3), (3,4)\}, \\

 \{(0,0), (1,1), (3,2), (4,5)\}, & \{(0,0), (1,2), (2,0), (5,2)\}, \\

 \{(0,0), (1,2), (2,1), (4,0)\}, & \{(0,0), (1,2), (3,1), (4,3)\}, \\

 \{(0,0), (1,2), (3,2), (5,1)\}, & \{(0,0), (1,3), (2,2), (3,5)\}, \\

 \{(0,0), (1,3), (3,1), (4,0)\}, & \{(0,0), (1,3), (3,3), (4,2)\}, \\

 \{(0,0), (1,4), (2,3), (4,5)\}, & \{(0,0), (1,4), (3,5), (5,1)\}. \\
\end{array}
\]
\qed

We finish this section by giving an optimal $(p,p,p+1,2)$-OOSPC from an inversive plane.

Let $q$ be a prime power and $ GF(q)$ the finite field of order $q$. Suppose that $a,b,c,d\in GF(q)$ and $ad-bc\not= 0$.
Define a linear fractional mapping $\pi_{{\footnotesize \left(\begin{array}{ll}
       a\ b\\
       c\ d\\
       \end{array}
       \right ) }
}:$  $GF(q)\cup
\{\infty\}\rightarrow GF(q)\cup
\{\infty\}$ as follows:
\[
\pi_{{\footnotesize \left (\begin{array}{ll}
       a\ b\\
       c\ d\\
       \end{array}
        \right ) }
}(x)=\left \{
\begin{array}{ll}
\frac{ax+b}{cx+d}, & {\rm if}\ x\in GF(q), cx+d\neq 0;\\
\infty, &           {\rm if}\ x\in GF(q), ax+b\neq 0, cx+d=0;\\
\frac{a}{c}, & {\rm if}\ x=\infty, c\neq 0;\\
\infty, &           {\rm if}\  x=\infty, c=0.
\end{array}
\right .
\]
Then $\pi_{{\footnotesize \left (\begin{array}{ll}
       a\ b\\
       c\ d\\
       \end{array}
        \right )} }$ is a permutation of $GF(q) \cup \{\infty\}$,
  and the permutations $\pi_{{\footnotesize \left (\begin{array}{ll}
       a\ b\\
       c\ d\\
       \end{array}
        \right )} }$ and $\pi_{{\footnotesize \left (\begin{array}{ll}
       ra\ rb\\
       rc\ rd\\
       \end{array}
        \right )} }$ are identical
if $r\neq  0$. Define PGL$(2, q)$ to consist of all the distinct permutations дл$\pi_{{\footnotesize \left (\begin{array}{ll}
       a\ b\\
       c\ d\\
       \end{array}
        \right )} }$,
where  $a,b,c,d\in GF(q)$, and $ad-bc\not= 0$. It is well known that
PGL$(2, q)$ is a sharply 3-transitive permutation group acting on the set $GF(q)\cup \{\infty\}$, i.e., for all choices of six elements $x_1, x_2, x_3$, $y_1, y_2, y_3 \in GF(q)\cup \{\infty\}$ such that $x_1, x_2, x_3$ are
distinct and $y_1, y_2, y_3$ are distinct, there is exactly one permutation $\pi\in$PGL$(2, q)$  such
that $\pi(x_i) = y_i$ for all $i$, $1 \leq i \leq 3$. Witt  proved
that the PGL$(2,q^2)$ orbit of the set $GF(q)\cup \{\infty\}$ is an $S(3,q+1,q^2+1)$ (called an {\em inversive plane}) with the point set $GF(q^2)\cup \{\infty\}$ \cite{Witt1938}.

\begin{theorem} {\rm \cite{Witt1938}}
For any prime power $q$, there is an $S(3,q+1,q^2+1)$.
\end{theorem}

\begin{theorem}
\label{(p,p,p+1,2)-OOSPC}
For any prime $p$, there is an optimal $(p,p,p+1,2)$-OOSPC with the size attaining the upper bound $(\ref{2D-JohnsonBound})$.
\end{theorem}

\proof  Start with an inversive plane $S(3,p+1,p^2+1)$ whose point set is $GF(p^2)\cup \{\infty\}$ and whose block set is the PGL$(2,p^2)$ orbit of the set $GF(p)\cup \{\infty\}$.
Since the PGL$(2,p^2)$ contains the permutation group $G=\{\pi_{{\footnotesize \left (\begin{array}{ll}
       1\ b\\
       0\ 1\\
       \end{array}
        \right )} }:b\in GF(p^2)\}$, the inversive plane $S(3,p+1,p^2+1)$ admits an automorphism group $G$. Since each automorphism $\pi_{{\footnotesize \left (\begin{array}{ll}
       1\ b\\
       0\ 1\\
       \end{array}
        \right )} }$ fixes the point $\infty$, the set of all blocks containing $\infty$ admits the automorphism group $G$. On the other hand, all blocks containing $\infty$ with $\infty$ deleted form the set of an $S(2,p,p^2)$ (called an affine plane). There are total $p^2+p$ blocks containing $\infty$.  Thus, deleting $\infty$ and all blocks containing $\infty$ from the inversive plane yields a $3$-$(p^2,p+1,1)$ packing design  which admits $G$ as a point-regular automorphism group and has $p^2(p-1)$ blocks. Since $gcd(p+1,p^2)=1$, the stabilizer of each block in the $3$-$(p^2,p+1,1)$ packing design under $G$ equals the identity of $G$, thus the $3$-$(p^2,p+1,1)$ packing design with point set $GF(p^2)$ is strictly $G$-invariant. This packing design is in fact a strictly $(GF(p^2),+)$-invariant $3$-$(p^2,p+1,1)$ packing design.
Since $(GF(p^2),+)$ is isomorphic to $\mathbb{Z}_p\times \mathbb{Z}_p$, there is a strictly $\mathbb{Z}_p\times \mathbb{Z}_p$-invariant  $3$-$(p^2,p+1,1)$ packing design with $p-1$ full orbits of blocks. By Theorem \ref{equivalence} and the upper bound (\ref{2D-JohnsonBound}), there is an optimal $(p,p,p+1,2)$-OOSPC. \qed

\section{Construction of strictly $\mathbb{Z}_m\times \mathbb{Z}_n$-invariant $G(m,n,4,3)$ designs via cyclic SQS$(m)$s}

In this section, we introduce a concept of strictly $\mathbb{Z}_m\times \mathbb{Z}_n$-invariant $G(m,n,4,3)$ design and present
a construction of a strictly $\mathbb{Z}_m\times \mathbb{Z}_n$-invariant PQS$(mn)$ from a strictly $\mathbb{Z}_m\times \mathbb{Z}_n$-invariant $G(m,n,4,3)$ design. We also use
a strictly semi-cyclic $G(2,n,4,3)$ design and a cyclic SQS$(m)$ to construct a strictly $\mathbb{Z}_m\times \mathbb{Z}_n$-invariant $G(m,n,4,3)$ design.

Let $m,n,t$ be positive integers and $K$ a set of some positive integers. A $G(m,n,K,t)$ design is a triple $(X,\Gamma,\B)$,
where $X$ is a set of $mn$ points, $\Gamma$ is a set of subsets of $X$ which is partition of $X$ into
$m$ sets of size $n$ (called {\it groups}) and $\B$ is a
set of subsets of $X$ with cardinalities from $K$, called {\it blocks},
such that each $t$-set of points not contained in any group occurs in exactly one block and each $t$-subset of each group does not occur in any block.
When $K=\{k\}$, we simply write $k$ for $K$.

$G$-designs were introduced by Mills \cite{Mills1974} who determined the existence of $G(m,6,4,3)$ design.  Recently, Zhuralev et al.
\cite{ZKK2008} showed that there exists a $G(m,n,4,3)$ design if and only if $n=1$ and
$m\equiv2,4\pmod{6}$, or $n$ is even and $n(m-1)(m-2)\equiv0
\pmod3$.

Let $m,n,t$ be positive integers and $K$ a set of some positive integers. An $H(m,n,K,t)$ design is a triple $(X,\Gamma,\B)$,
where $X$ is a set of $mn$ points, $\Gamma$ is a partition of $X$ into
$m$ sets of size $n$ (called {\it groups}) and $\B$ is a
set of subsets of $X$ with cardinalities from $K$, called {\it blocks},
such that each block intersect each group in at most one point and each $t$-set of points from $t$ distinct groups  occurs in exactly one block.

The early idea of an $H$-design can be found in Hanani
\cite{Hanani1963}, who used different terminology. Mills used the
terminology $H$-design in \cite{Mills1974} and
determined the existence of an $H(m,n,4,3)$ design except
for $m=5$ \cite{mills1990}. Recently, the second author proved that an $H(5,n,4,3)$ design exists if $n$ is even,
$n\neq2$ and $n\not\equiv 10,26\pmod{48}$  \cite{Ji2009} .

An {\em automorphism} $\alpha$ of a $G$-design (resp. $H$-design) $(X,\Gamma,{\cal B})$ is
a permutation on $X$ leaving $\Gamma$ and ${\cal B}$ invariant.
All  automorphisms of an $G$-design (resp. $H$-design) form a group, called the {\em full automorphism group} of the  $G$-design (resp. $H$-design).
Any subgroup of the full automorphism group is called an {\em automorphism group} of the $G$-design (resp. $H$-design).  A $G$-design (resp. $H$-design) $(X,\Gamma,\B)$ is said to be $Q$-{\em invariant} if it admits $Q$ as a point-regular automorphism group. Moreover, it
is said to be {\em strictly $Q$-invariant} if it is $Q$-invariant and the stabilizer of each $B\in \B$ under $Q$ equals the identity of $Q$.
For  a  $Q$-invariant
$G$-design (resp. $H$-design), we always identify
the point set $X$ with $Q$ and  the automorphisms are
regarded as translations $\sigma_a$ defined by $\sigma_a(x)= x+a$ for $x\in Q$, where $a\in Q$. Then all blocks of this $G$-design (resp. $H$-design) can be partitioned into some orbits under the permutation group $\{\sigma_a:a\in Q\}$.
Let $L$ be a subgroup of $Q$.
If the group set of a $Q$-{\em invariant}  $G$-design (resp. $H$-design) is a set of cosets of $L$ in $Q$, then it is a $Q$-{\em invariant}  $G$-design (resp. $H$-design) relative to $L$.
A $G(m,n,K,t)$ (resp. $H(m,n,K,t)$) design is said to be {\em semi-cyclic} if the $G$-design (resp. $H$-design) admits an automorphism $\sigma$ consisting of $m$ cycles of length $n$ and leaving each group invariant. Note that the stabilizer of each block $B$ of a semi-cyclic $H$-design under $\{\sigma^i: 0\leq i<n\}$ equals the identity, i.e., a semi-cyclic $H$-design is always strictly.

\begin{example}
\label{Z10Z2SQS(20)}
There is a strictly $\mathbb{Z}_{10}\times \mathbb{Z}_2$-invariant $G(5,4,4,3)$ design relative to $5\mathbb{Z}_{10}\times \mathbb{Z}_2$.
\end{example}

\proof The following base blocks  under $\mathbb{Z}_{10}\times \mathbb{Z}_2$ generate the set of blocks of a strictly $\mathbb{Z}_{10}\times \mathbb{Z}_2$-invariant $G(5,4,4,3)$  design over $\mathbb{Z}_{10}\times \mathbb{Z}_2$ with groups $\{i,i+5\}\times \mathbb{Z}_2$, $0\leq i<5$.
\[
\begin{array}{llll}
\{(0,0),(1,0),(9,0),(0,1)\}, & \{(0,0),(2,0),(8,0),(0,1)\}, \\ \{(0,0),(3,0),(7,0),(0,1)\}, &
\{(0,0),(4,0),(6,0),(0,1)\}, \\  \{(0,0),(1,0),(3,0),(4,0)\}, & \{(0,0),(1,0),(5,0),(6,1)\}, \\
\{(0,0),(1,0),(6,0),(5,1)\}, & \{(0,0),(1,0),(2,1),(3,1)\}, \\ \{(0,0),(1,0),(4,1),(7,1)\}, &
\{(0,0),(2,0),(5,0),(7,1)\}, \\  \{(0,0),(2,0),(7,0),(5,1)\}, & \{(0,0),(2,0),(3,1),(9,1)\}, \\
 \{(0,0),(2,0),(4,1),(8,1)\}, &  \{(0,0),(3,0),(1,1),(4,1)\}.\\
\end{array}
\] \qed

%\begin{example}
%\label{Z8Z4-G(4,8,4,3)}
%There exists a strictly $\mathbb{Z}_8\times \mathbb{Z}_4$-invariant $G(4,8,4,3)$ over  $\mathbb{Z}_8\times \mathbb{Z}_4$ with group set $\{\{i,i+4\}\times \mathbb{Z}_4:0\leq i<4\}$.
%\end{example}
%\proof The following base blocks generate the block set  of the required strictly $\mathbb{Z}_8\times \mathbb{Z}_4$-invariant $G(4,8,4,3)$.
%\[
%\begin{array}{ll}
%\{(0,0),(1,0),(2,0),(4,0)\}, & \{(0,0),(1,0),(5,0),(0,1)\}, \\ \{(0,0),(1,0),(6,0),(2,1)\}, & \{(0,0),(1,0),(1,1),(3,1)\}, \\
%
%\{(0,0),(1,0),(4,1),(5,1)\}, & \{(0,0),(1,0),(6,1),(7,1)\}, \\ \{(0,0),(1,0),(0,2),(2,2)\}, & \{(0,0),(1,0),(1,2),(4,2)\},  \\
%
%\{(0,0),(1,0),(3,2),(5,2)\}, & \{(0,0),(1,0),(6,2),(0,3)\}, \\ \{(0,0),(1,0),(7,2),(6,3)\}, & \{(0,0),(1,0),(1,3),(7,3)\},  \\
%
%\{(0,0),(2,0),(5,0),(7,2)\}, & \{(0,0),(2,0),(0,1),(1,3)\}, \\ \{(0,0),(2,0),(3,1),(2,2)\}, & \{(0,0),(2,0),(5,1),(2,3)\},  \\
%
%\{(0,0),(2,0),(6,1),(4,3)\}, & \{(0,0),(2,0),(7,1),(3,3)\}, \\ \{(0,0),(2,0),(3,2),(5,3)\}, & \{(0,0),(2,0),(4,2),(6,3)\},  \\
%
%\{(0,0),(3,0),(0,1),(7,3)\}, & \{(0,0),(3,0),(1,1),(3,3)\}, \\ \{(0,0),(3,0),(2,1),(4,3)\}, & \{(0,0),(3,0),(5,1),(5,3)\},  \\
%
%\{(0,0),(3,0),(6,1),(6,2)\}, & \{(0,0),(3,0),(7,1),(6,3)\}, \\ \{(0,0),(3,0),(1,2),(1,3)\}, & \{(0,0),(3,0),(3,2),(0,3)\},  \\
%
%\{(0,0),(3,0),(4,2),(2,3)\}, & \{(0,0),(4,0),(1,1),(6,2)\}, \\ \{(0,0),(4,0),(2,1),(5,3)\}, & \{(0,0),(4,0),(1,2),(6,3)\},  \\
%
%\{(0,0),(4,0),(3,2),(3,3)\}, & \{(0,0),(1,1),(2,2),(5,3)\}, \\ \{(0,0),(1,1),(3,2),(1,3)\}, & \{(0,0),(1,1),(5,2),(4,3)\},  \\
%
%\{(0,0),(3,1),(1,2),(4,3)\}.
%\end{array}
%\]
%\qed

For  a semi-cyclic $G(m,n,K,t)$ (resp. $H(m,n,K,t)$)  design,  without loss of generality we can identify
the point set $X$ with $I_m\times \mathbb{Z}_n$, and the automorphism $\sigma$
can be taken as $(i,j)\mapsto (i,j+1)$ $(-,{\rm mod}\ n)$,  $(i,j)\in I_m\times \mathbb{Z}_n$, where $I_m=\{1,\ldots,m\}$. Then all blocks of this $G$-design (resp. $H$-design) can be partitioned into some orbits under the action of $\sigma$. A set of base blocks is a set of representatives for the orbits and each element is  called a base block.

Let $n$ be a positive integer. It is not hard to see that $$\{\{(1,x),(2,x+y),(3,x+z),(4,x+y+z)\}:\ x,y,z\in \mathbb{Z}_n\}$$ is the set of a semi-cyclic $H(4,n,4,3)$  design on $I_4\times \mathbb{Z}_n$ with groups $\{i\}\times \mathbb{Z}_n$, $i\in I_4$. Such a result has been stated in \cite{FC2011}.

\begin{lemma}
\label{semi-cyclic H(6n+2,m,4,3)}
{\rm \cite{FC2011}} For any positive integers $n$, there exists a semi-cyclic $H(4, n, 4, 3)$  design.
\end{lemma}

The following construction is simple but very useful.

\begin{construction}
\label{filling} Let $e,m,n$ be positive integers  such that $m$ is divisible by $e$. Suppose there exists a strictly $\mathbb{Z}_{m}\times \mathbb{Z}_n$-invariant
$G(\frac{m}{e},en,k,3)$  design relative to $\frac{m}{e}\mathbb{Z}_{m}\times \mathbb{Z}_n$. If there exists a strictly $\mathbb{Z}_e\times \mathbb{Z}_n$-invariant $3$-$(en,k,1)$ packing design having $b$ base blocks,
then there exists a strictly $\mathbb{Z}_{m}\times \mathbb{Z}_n$-invariant $3$-$(mn,k,1)$ packing design having $b+\frac{(mn-1)(mn-2)-(en-1)(en-2)}{k(k-1)(k-2)}$ base blocks. Further, if $b=J(e,n,k,2)$ then $b+\frac{(mn-1)(mn-2)-(en-1)(en-2)}{k(k-1)(k-2)}$ $=J(m,n,k,2)$.
\end{construction}

\proof  Let ${\cal F}$ be the family of base blocks of a strictly $\mathbb{Z}_{m}\times \mathbb{Z}_n$-invariant
$G(\frac{m}{e},en,k,3)$  design with group set $\{\{i,i+\frac{m}{e},\ldots, i+m-\frac{m}{e}\}\times \mathbb{Z}_n:0\leq i<\frac{m}{e}\}$. Then $|{\cal F}|=\frac{(mn-1)(mn-2)-(en-1)(en-2)}{k(k-1)(k-2)}$. Let ${\cal A}$ be
the family of $b$ base blocks of a strictly $(\frac{m}{e}\mathbb{Z}_m)\times \mathbb{Z}_n$-invariant $3$-$(en,k,1)$ packing design, where $\frac{m}{e}\mathbb{Z}_m=\{0,\frac{m}{e},\frac{2m}{e},\ldots,m-\frac{m}{e}\}$. Such a design exists by assumption.
Then ${\cal F}\bigcup {\cal A}$
is  the set of base blocks of a strictly $\mathbb{Z}_{m}\times \mathbb{Z}_n$-invariant $3$-$(mn,k,1)$ packing design having $b+\frac{(mn-1)(mn-2)-(en-1)(en-2)}{k(k-1)(k-2)}$ base blocks.

When $b=J(e,n,k,2)$, the same discussion as the proof of \cite[Theorem 6.6]{FCJ2008} shows that
\begin{equation*}
\label{3}\frac{(mn-1)(mn-2)-(en-1)(en-2)}{k(k-1)(k-2)}+
\left \lfloor\frac{1}{k}\left \lfloor\frac{en-1}{k-1}
\left \lfloor\frac{en-2}{k-2}\right \rfloor\right\rfloor\right\rfloor=\left \lfloor\frac{1}{k}\left \lfloor\frac{mn-1}{k-1}
\left \lfloor\frac{mn-2}{k-2}\right\rfloor\right\rfloor\right\rfloor.
\end{equation*}

For completeness, we give its proof again.
First we will prove that if $a$, $b$, $c$ are positive integers,
then $\lfloor\frac{1}{a}\lfloor\frac{c}{b}\rfloor\rfloor
=\lfloor\frac{c}{ab}\rfloor.$ Let $c=xb+y$, $0\leq y\leq b-1$. Let
$x=x_1 a+y_1$, $0 \leq y_1 \leq a-1$. It follows that
$\lfloor\frac{c}{ab}\rfloor=\lfloor\frac{x}{a}+\frac{y}{ab}\rfloor=\lfloor
x_1+\frac{y_1}{a}+\frac{y}{ab}\rfloor=\lfloor x_1+\frac{y_1
b+y}{ab}\rfloor=x_1=\lfloor\frac{1}{a}\lfloor\frac{c}{b}\rfloor\rfloor$.

For any given pair $P$ of points from a group, consider all triples containing $P$.  By the definition of a $G$-design, there are total $mn-en$ such triples and each block containing $P$ contains $k-2$ such triples. Therefore, $mn-en$ is divisible by $k-2$. Similarly, for any given pair $P$ of points from distinct groups, consider all triples containing $P$, we then obtain that $mn-2$ is divisible by $k-2$, thereby, $en-2$ is also divisible by $k-2$.
Then $\frac{(mn-1)(mn-2)-(en-1)(en-2)}{k(k-1)(k-2)}+
\left \lfloor\frac{1}{k}\left \lfloor\frac{en-1}{k-1}
\left \lfloor\frac{en-2}{k-2}\right \rfloor\right\rfloor\right\rfloor$ is equal to
\begin{displaymath}
\begin{array}{l}
\frac{(mn-1)(mn-2)-(en-1)(en-2)}{k(k-1)(k-2)}+\left \lfloor\frac{(en-1)(en-2)}{k(k-1)(k-2)}
\right \rfloor=\left \lfloor\frac{(mn-1)(mn-2)}{k(k-1)(k-2)}
\right \rfloor
=\left \lfloor\frac{1}{k}\left \lfloor\frac{mn-1}{k-1}
\left \lfloor\frac{mn-2}{k-2}\right\rfloor\right\rfloor\right\rfloor,
\end{array}
\end{displaymath}
as desired. \qed

Construction \ref{filling} shows that it is useful to find some
strictly $\mathbb{Z}_{m}\times \mathbb{Z}_n$-invariant $G(\frac{m}{e},en,4,3)$  designs.

A block-orbit of a cyclic SQS$(v)$ is said to be quarter if the
block-orbit contains the block $\{0,v/4,v/2,3v/4\}$,
while a block-orbit of a cyclic SQS$(v)$ is said to be half if the
block-orbit contains a block of the form $\{0,i,v/2,v/2+i\}$, $0<i<v/4$.
It is easy to see that in a  cyclic SQS$(v)$, each block-orbit is full, half, or quarter.

\begin{construction}
\label{CSQS(m)-ZmZnG(m/2,2n,4,3)}
If there is a cyclic SQS$(m)$ and a strictly semi-cyclic $G(2,n,4,3)$  design with $n>1$, then there is a strictly $\mathbb{Z}_m\times \mathbb{Z}_n$-invariant $G(\frac{m}{2},2n,4,3)$  design relative to $\frac{m}{2}\mathbb{Z}_{m}\times \mathbb{Z}_n$.
\end{construction}

\proof By the necessary condition of a $G(2,n,4,3)$  design, we have that $n$ is even. First, we construct a  strictly $\mathbb{Z}_m\times \mathbb{Z}_2$-invariant
$H(m,2,4,3)$  design over $\mathbb{Z}_m\times \mathbb{Z}_2$ with the group set $\{\{i\}\times \mathbb{Z}_2:i\in \mathbb{Z}_m\}$.

Let $(\mathbb{Z}_m,\B)$ be a cyclic SQS$(m)$. Let $\widetilde{\B}_1$ be the set of base blocks generating all full block-orbits, $\widetilde{\B}_2$ the set of base blocks generating all half block-orbits, $\widetilde{\B}_3$ the set of the base block generating the unique quarter block-orbit.  Note that $\widetilde{\B}_2, \widetilde{\B}_3$ may be empty sets.
% Then $\frac{m(m-1)(m-2)}{24}=m\cdot |\widetilde{\B}_1|+\frac{m}{2}\cdot |\widetilde{\B}_2|+\frac{v}{4}|\cdot \widetilde{\B}_3|$.

Take any base block $B=\{x,y,z,w\}\in \widetilde{\B}_1\cup \widetilde{\B}_2\cup \widetilde{\B}_3$, construct a semi-cyclic $H(4,2,4,3)$  design on $B\times \mathbb{Z}_2$ with the group set $\{\{x\}\times \mathbb{Z}_2:x\in B\}$  and the following  eight blocks:
\[
\begin{array}{c}
 \{(x,0),(y,0),(z,0),(w,1)\}, \ \{(x,1),(y,1),(z,1),(w,0)\},\\
 \{(x,0),(y,0),(z,1),(w,0)\}, \ \{(x,1),(y,1),(z,0),(w,1)\},\\
 \{(x,0),(y,1),(z,0),(w,0)\}, \ \{(x,1),(y,0),(z,1),(w,1)\},\\
 \{(x,1),(y,0),(z,0),(w,0)\}, \ \{(x,0),(y,1),(z,1),(w,1)\}.\\
\end{array}
\]
Clearly, the  four blocks on the right are obtained by adding $(0,1)$ to the four blocks on the left.
When $B\in \widetilde{\B}_2$, it must be of the form $\{0,i,m/2,i+m/2\}+j$. Let $x=j,y=j+i,z=j+m/2$ and $w=j+i+m/2$.
Then it is easy to see that the later four blocks are obtained by adding $(m/2,0)$ to the first four blocks, respectively.  When $B\in \widetilde{\B}_3$, it must be of the form $\{0,m/4,m/2,3m/4\}+j$. Let $x=j,y=j+m/4,z=j+m/2$ and $w=j+3m/4$. Then it is easy to see that the later six blocks are obtained by adding $(m/4,0),(m/2,0),(3m/4,0)$ to the first two blocks, respectively. Let $\A_B^1$ consist of the four blocks on the left if $B\in  \widetilde{\B}_1$, let $\A_B^2$ consist of the first two blocks on the left if $B\in  \widetilde{\B}_2$, and let $\A_B^3$ consist of the first block if $B\in  \widetilde{\B}_3$.

Denote $${\cal A}=(\bigcup_{B\in \widetilde{\B}_1}\A_B^1)\bigcup (\bigcup_{B\in \widetilde{\B}_2}\A_B^2)\bigcup (\bigcup_{B\in \widetilde{\B}_3}\A_B^3).$$
It is routine to check that $\A$ is the set of base blocks of the required
strictly $\mathbb{Z}_m\times \mathbb{Z}_2$-invariant
$H(m,2,4,3)$  design.

Secondly, we construct a strictly $\mathbb{Z}_m\times \mathbb{Z}_n$-invariant
$H(m,n,4,3)$  design over $\mathbb{Z}_m\times \mathbb{Z}_n$ with the group set $\{\{i\}\times \mathbb{Z}_n:i\in \mathbb{Z}_m\}$.

For $n>2$, write $n=2n'$. For each base block $A\in \A$, construct a semi-cyclic
$H(4,n',4,3)$  design on $A\times \mathbb{Z}_{n'}$ with groups $\{\{x\}\times \mathbb{Z}_{n'}: x\in
A\}$. Such a design exists by Lemma \ref{semi-cyclic H(6n+2,m,4,3)}. Denote the family
of base blocks of this design by $\C_A$. Define a mapping $\varphi:$ $\mathbb{Z}_{m}\times \mathbb{Z}_2\times \mathbb{Z}_{n'}\rightarrow\mathbb{Z}_{m}\times \mathbb{Z}_{n}$ by $\varphi(i,\ell,k)=(i,\ell+2k)$ for $(i,\ell,k)\in \mathbb{Z}_{m}\times \mathbb{Z}_2\times \mathbb{Z}_{n'}$.
Denote $\D=\{\{\varphi(z):z\in C\}: C\in \C_A, A\in \A\}$ and let $\D'=\{D+\delta:\ D\in \D,\ \delta\in \mathbb{Z}_{m}\times \mathbb{Z}_n\}$.

Simple computation shows that $|\D|=|{\cal A}|\cdot |\C_A|=\frac{(m-1)(m-2)}{6}|\C_A|=(n')^2\cdot \frac{(m-1)(m-2)}{6}=\frac{(m-1)(m-2)n^2}{24}$, which is the right number of base blocks of a strictly $\mathbb{Z}_m\times \mathbb{Z}_n$-invariant $H(m,n,4,3)$  design. We need only to show that each triple from three distinct groups appears in at least one block of $\D'$.  Let $T=\{(i_1,\ell_1+2k_1),(i_2,\ell_2+2k_2),(i_3,\ell_3+2k_3)\}$ be such a triple, where $i_1,i_2,i_3$ are distinct, $\ell_j\in \{0,1\}$ and $k_j\in \{0,1,\ldots,n'-1\}$ for $j\in \{1,2,3\}$.

Since $\{A+\tau:A\in \A, \tau\in \mathbb{Z}_m\times \mathbb{Z}_2\}$ is a strictly $\mathbb{Z}_m\times \mathbb{Z}_2$-invariant $H(m,2,4,3)$  design, there is a base block $A=\{(a_1,c_1),(a_2,c_2),(a_3,c_3),(a_4,c_4)\}\in \A$ and an element $(\delta_1,\delta_2)$, $0\leq \delta_1<m$ and $\delta_2\in \{0,1\}$, such that $\{(i_1,\ell_1),(i_2,\ell_2),(i_3,\ell_3)\}\subset \{(a_1,c_1),(a_2,c_2),(a_3,c_3),(a_4,c_4)\}+(\delta_1,\delta_2)$. Without loss of generality, let $a_j+\delta_1\equiv i_j\pmod m$ and $\ell_j\equiv c_j+\delta_2\pmod 2$.
Denote $ c_j+\delta_2=\ell_j+2\sigma_j$, $\sigma_j\in \{0,1\}$. Since $\C_A$ is the set of base blocks of a semi-cyclic $H(4,n',4,3)$  design over $A\times \mathbb{Z}_{n'}$, there is a base block $C=\{(a_1,c_1,d_1),(a_2,c_2,d_2),(a_3,c_3,d_3)$,$(a_4,c_4,d_4)\}\in \C_A$ and an element $\delta_3\in \{0,1,\ldots,n'-1\}$ such that $k_j-\sigma_j\equiv d_j+\delta_3\pmod {n'}$. It follows that $T\subset \varphi(C)+(\delta_1,\delta_2+2\delta_3)\in \D'$.

Finally, we construct a strictly $\mathbb{Z}_m\times \mathbb{Z}_n$-invariant $G(\frac{m}{2},2n,4,3)$  design relative to $\frac{m}{2}\mathbb{Z}_{m}\times \mathbb{Z}_n$.

For $1\leq i<\frac{m}{2}$, construct a strictly semi-cyclic $G(2,n,4,3)$  design on $\{0,i\}\times \mathbb{Z}_n$ with groups $\{0\}\times \mathbb{Z}_n$ and $\{i\}\times \mathbb{Z}_n$. Such a design exists by assumption. Denote the set of base blocks by $\F_i$ and let $\F=\cup_{1\leq i<m/2}\F_i$.
It is easy to see that $\D\cup \F$ is the set of base blocks of the required strictly $\mathbb{Z}_m\times \mathbb{Z}_n$-invariant $G(\frac{m}{2},2n,4,3)$  design.
\qed

%\begin{center}
%\begin{tabular}[t]{|p{13cm}|}
%\hline
% {\textbf step1} construction construction construction construction construction \\
% \\
% step2 construction construction construction construction construction \\
% step3 construction construction  construction  ~~~ construction construction \\
%\hline
%\end{tabular}
%
%\ Example of
%\end{center}

We illustrate the idea of Construction \ref{CSQS(m)-ZmZnG(m/2,2n,4,3)} with $m=4$ and $n=8$.
\begin{example}
\label{Z4Z8G(2,16,4,3)}
There is a strictly $\mathbb{Z}_4\times \mathbb{Z}_8$-invariant $G(2,16,4,3)$  design relative to $2\mathbb{Z}_{4}\times \mathbb{Z}_8$
and an optimal $(4,8,4,2)$-OOSPC with the size meeting the upper bound $(\ref{2D-JohnsonBound})$.
\end{example}

\begin{itemize}
\item Step 1:  Since the trivial cyclic SQS$(4)$ has a unique block $\{0,1,2,3\}$, we have $A=\{(0,0),(1,0)$, $(2,0),(3,1)\}$ is the unique base block of a strictly $\mathbb{Z}_4\times \mathbb{Z}_2$-invariant $H(4,2,4,3)$  design with groups  $\{i\}\times \mathbb{Z}_2$, $i\in \mathbb{Z}_4$, i.e., $\A=\{A\}$.

\item Step 2:  Since $\{\{(0,0,0),(1,0,x),(2,0,y),(3,1,x+y)\}\colon 0\leq x,y<4\}$ is the set of base blocks of  a  semi-cyclic $H(4,4,4,3)$  design on $A\times \mathbb{Z}_4$ with groups $\{z\}\times \mathbb{Z}_4$ ($z\in A$), by the mapping $\varphi:$ $\mathbb{Z}_{4}\times \mathbb{Z}_2\times \mathbb{Z}_{4}\rightarrow\mathbb{Z}_{4}\times \mathbb{Z}_{8}$ defined by $\varphi(i,\ell,k)=(i,\ell+2k)$ for $(i,\ell,k)\in \mathbb{Z}_{4}\times \mathbb{Z}_2\times \mathbb{Z}_{4}$ we have
    $$\D=\{\{(0,0),(1,2x),(2,2y),(3,1+2x+2y)\}\colon 0\leq x,y<4\}$$ is the set of base blocks of a strictly $\mathbb{Z}_4\times \mathbb{Z}_8$-invariant $H(4,8,4,3)$  design with groups  $\{i\}\times \mathbb{Z}_8$, $i\in \mathbb{Z}_4$.
\item Step 3: Construct a strictly semi-cyclic $G(2,8,4,3)$  design on $\{0,1\}\times \mathbb{Z}_8$ with groups $\{i\}\times \mathbb{Z}_8$, $i\in \{0,1\}$, whose set $\F$ of base blocks consists of the following:
\[
 \begin{array}{ll}
 \{(0,0), (0,1), (1,0), (1,1)\}, & \{(0,0), (0,1), (1,2), (1,4)\}, \\
 \{(0,0), (0,1), (1,3), (1,7)\}, & \{(0,0), (0,1), (1,5), (1,6)\}, \\
 \{(0,0), (0,2), (1,0), (1,2)\}, & \{(0,0), (0,2), (1,1), (1,6)\}, \\
\{(0,0), (0,2), (1,3), (1,4)\}, & \{(0,0), (0,2), (1,5), (1,7)\}, \\
\{(0,0), (0,3), (1,0), (1,4)\}, & \{(0,0), (0,3), (1,1), (1,7)\}, \\
\{(0,0), (0,3), (1,2), (1,5)\}, & \{(0,0), (0,3), (1,3), (1,6)\}, \\
\{(0,0), (0,4), (1,0), (1,5)\}, & \{(0,0), (0,4), (1,2), (1,3)\}. \\
 \end{array}
\]
Then  $\D\cup \F$ is the set of base blocks of a strictly $\mathbb{Z}_4\times \mathbb{Z}_8$-invariant $G(2,16,4,3)$  design with groups  $\{i,i+2\}\times \mathbb{Z}_8$, $0\leq i<2$.
\end{itemize}

Since there is a $(2,8,4,2)$-OOSPC with $J(2,8,4,2)$ codewords from \cite{Sawa2010},  there is a strictly $\mathbb{Z}_2\times \mathbb{Z}_8$-invariant $PQS(16)$ with $J(2,8,4,2)$ base blocks by Theorem \ref{equivalence}. By Construction \ref{filling}, there is a strictly $\mathbb{Z}_4\times \mathbb{Z}_8$-invariant $PQS(32)$ with  $J(4,8,4,2)$ base blocks, which leads to an optimal $(4,8,4,2)$-OOSPC with the size meeting the upper bound $(\ref{2D-JohnsonBound})$ by  Theorem \ref{equivalence}. \qed

\section{Constructions of strictly $\mathbb{Z}_m\times \mathbb{Z}_n$-invariant $G^*(m,n,4,3)$  designs}

In this section, we give product constructions of strictly $\mathbb{Z}_m\times \mathbb{Z}_n$-invariant $G^*(m,n,4,3)$  designs.

Let $e,m,n$ be positive integers such that $m$ is divisible by $e$.
Let $m-e\equiv n\equiv 0\pmod 2$. In a $\mathbb{Z}_{m}\times \mathbb{Z}_n$-invariant $G(\frac{m}{e},en,4,3)$  design $(\mathbb{Z}_{m}\times \mathbb{Z}_n,\{\{i,i+\frac{m}{e},\ldots,m-\frac{m}{e}\}\times \mathbb{Z}_n: 0\leq i<\frac{m}{e}\},{\cal B})$,
there exist $n(m-e)/2$ triples of
the form $\{(0,0),(i,j),(-i,-j)\}$, $(i,j)\in \mathbf{I}\times \mathbb{Z}_n$, and $(m-e)n$ triples of the form $\{(0,0),(i,j),(0,n/2)\}$, $(i,j)\in (\mathbb{Z}_{m}\setminus \frac{m}{e}\mathbb{Z}_{m})\times \mathbb{Z}_n$, respectively, where $\mathbf{I}=\{k: 1\leq k\leq \lfloor\frac{m}{2}\rfloor,k\not\equiv 0\pmod {\frac{m}{e}}\}$ and $\frac{m}{e}\mathbb{Z}_{m}=\{0,\frac{m}{e},\ldots,m-\frac{m}{e}\}$. If any triple of the form $\{y,y+x,y-x\}$ or
$\{y,y+z,y+(0,n/2)\}$, where $x\in \mathbf{I}\times \mathbb{Z}_n$, $z\in  (\mathbb{Z}_{m}\setminus \frac{m}{e}\mathbb{Z}_{m})\times \mathbb{Z}_n$ and $y\in \mathbb{Z}_m\times \mathbb{Z}_n$, is contained in the block
$\{y,y+a,y-a,y+(0, n/2)\}$ for some $a\in
\mathbf{I}\times \{0,1,\ldots, \frac{n}{2}-1\}$, then such a
$G(\frac{m}{e},en,4,3)$  design is denoted by $G^*(\frac{m}{e},en,4,3)$.

In Example \ref{Z10Z2SQS(20)}, the first four base blocks generate 80 blocks which contain all triples of the form $\{y,y+x,y-x\}$,
$\{y,y+z,y+(0,1)\}$, where $x\in \{1,2,3,4\}\times \mathbb{Z}_2$, $z\in  (\mathbb{Z}_{10}\setminus \{0,5\})\times \mathbb{Z}_2$ and $y\in \mathbb{Z}_{10}\times \mathbb{Z}_2$, thereby, this $G$-design is also a  strictly $\mathbb{Z}_{10}\times \mathbb{Z}_2$-invariant $G^*(5,4,4,3)$.

Two constructions for $\mathbb{Z}_{m}\times \mathbb{Z}_n$-invariant $G^*$-designs are presented in Constructions \ref{G*-design} and \ref{G*-design-1}. The proofs of constructions are of design theory. Here, we only describe how to construct them. The detailed proof of  Construction \ref{G*-design} is moved to Appendix A. The detailed proof of Construction \ref{G*-design-1} is omitted, which is similar to that of Construction \ref{G*-design}.

\begin{construction}
\label{G*-design} Let $m,n,e,g$ be positive integers such that $m$ is divisible by $e$, both $n$ and $m-e$ are even, $g$ is odd and $g\geq 3$. If there
exists a strictly $\mathbb{Z}_{m}\times \mathbb{Z}_n$-invariant $G^*(\frac{m}{e},en,4,3)$  design relative to $\frac{m}{e}\mathbb{Z}_m\times \mathbb{Z}_n$, then
there exists a strictly $\mathbb{Z}_{m}\times \mathbb{Z}_{ng}$-invariant $G^*(\frac{m}{e},eng,4,3)$  design relative to $\frac{m}{e}\mathbb{Z}_m\times \mathbb{Z}_{ng}$.
\end{construction}

\proof  Let $(\mathbb{Z}_{m}\times \mathbb{Z}_n,\{\{i,i+\frac{m}{e},\ldots,i+m-\frac{m}{e}\}\times \mathbb{Z}_n:0\leq i<\frac{m}{e}\},{\cal B})$ be a strictly $\mathbb{Z}_{m}\times \mathbb{Z}_n$-invariant $G^*(\frac{m}{e},en,4,3)$  design. Let $\mathbf{I}=\{i:1\leq i\leq \lfloor\frac{m}{2}\rfloor, i\not\equiv 0\pmod {\frac{m}{e}}\}$. Denote the
family of base blocks of this design by ${\cal F}={\cal F}_1\bigcup
{\cal F}_2$, where ${\cal F}_1$ consists of all base blocks in the
form of $\{(0,0),(0,\frac{n}{2}),(i,j),(-i,-j)\}$ for $i\in \mathbf{I}$ and $0\leq j<\frac{n}{2}$, and ${\cal
F}_2$ consists of all the other base blocks. It is easy to see that
$|\F_1|=n(m-e)/4$ and $|\F_2|=n(m-e)(mn+en-9)/24$.
We construct the required $\mathbb{Z}_{m}\times \mathbb{Z}_{ng}$-invariant $G^*(\frac{m}{e},eng,4,3)$  design on $\mathbb{Z}_{m}\times \mathbb{Z}_{ng}$ with
group set $\G=\{\{i,i+\frac{m}{e},\ldots,i+m-\frac{m}{e}\}\times \mathbb{Z}_{ng}:\ 0\leq i<\frac{m}{e}\}$.

Define
\[
\begin{array}{l}
\vspace{2pt} \C_1=\{\{(i_0,j_0),(i_1,j_1+k_1n),(i_2,j_2+k_2n),(i_3,j_3+k_1n+k_2n)\}\colon \{(i_0,j_0),\ldots,(i_3,j_3)\}\in \F_2,\\
\hspace{10cm} 0\leq k_1,k_2< g\}, \\
\vspace{2pt} \C_2=\{\{(0,0),(i,j'),(-i,-j'),(0,\frac{ng}{2})\}\colon i\in \mathbf{I}, 0\leq j'<\frac{ng}{2}\},\\
\vspace{2pt} \C_3=\{\{(0,0),(i,j+\ell n),(i,j+\ell' n), (2i,2j+\ell n+\ell' n)\}\colon i\in \mathbf{I},\ 0\leq
j< n,\ 0\leq \ell<\ell'<g\},\\
\vspace{2pt} \C_4=\{\{(0,0),(0,\frac{n}{2}+\ell n), (i,j+\ell'n),(i,j+\frac{n}{2}+\ell n+\ell'n)\}\colon i\in \mathbf{I},\ 0\leq
j< \frac{n}{2},\ 0\leq \ell<\frac{g-1}{2},\\ \hspace{9.2cm} 0\leq \ell'<g\}.\\
\end{array}
\]
Note that for each base block $B=\{(i_0,j_0),(i_1,j_1),(i_2,j_2),(i_3,j_3)\}\in \F_2$, ${\cal A}_B=\{\{(i_0,j_0),(i_1,j_1+k_1n),(i_2,j_2+k_2n),(i_3,j_3+k_1n+k_2n)\}\colon 0\leq k_1,k_2< g\}$ is the set of base blocks of  a semi-cyclic
$H(4,g,4,3)$  design on $\{(x,y+kn):(x,y)\in B, 0\leq k<g\}$ with group set $\{\{(x,y+kn):0\leq k<g\}: (x,y)\in
B\}$ through $+(0,n)$ mod $(m,ng)$.

Let $\C_i'=\{C+\delta:\ C\in \C_i,\ \delta\in \mathbb{Z}_{m}\times \mathbb{Z}_{ng}\}$ for $1\leq i\leq 4$.
Denote $\C'=\C_1'\cup \C_2'\cup \C_3'\cup \C_4'$. We claim that $\C'$ is the set of blocks of the required strictly $\mathbb{Z}_{m}\times \mathbb{Z}_{ng}$-invariant $G^*(\frac{m}{e},eng,4,3)$  design. \qed

\begin{construction}
\label{G*-design-1} Let $m,n,e,g$ be positive integers such that $m$ is divisible by $e$, both $n$ and $m-e$ are even, $g$ is odd and $g\geq 3$. If there
exists a strictly $\mathbb{Z}_{m}\times \mathbb{Z}_n$-invariant $G^*(\frac{m}{e},en,4,3)$  design relative to $\frac{m}{e}\mathbb{Z}_m\times \mathbb{Z}_n$, then
there exists a strictly $\mathbb{Z}_{mg}\times \mathbb{Z}_{n}$-invariant $G^*(\frac{m}{e},egn,4,3)$  design relative to $\frac{m}{e}\mathbb{Z}_{mg}\times \mathbb{Z}_n$.
\end{construction}

\proof  We keep the notations of Construction \ref{G*-design} and we adapt the proof to the present
situation. Define
\[
\begin{array}{l}
\vspace{2pt} \D_1=\{\{(i_0,j_0),(i_1+k_1m,j_1),(i_2+k_2m,j_2),(i_3+k_1m+k_2m,j_3)\}\colon \{(i_0,j_0),\ldots,(i_3,j_3)\}\in \F_2,\\
\hspace{10.5cm} 0\leq k_1,k_2< g\}, \\
\vspace{2pt} \D_2=\{\{(0,0),(i+\ell m,j),(-i-\ell m,-j),(0,\frac{n}{2})\}\colon i\in \mathbf{I}, 0\leq j<\frac{n}{2},\ 0\leq \ell<g\},\\
\vspace{2pt} \D_3=\{\{(0,0),(i+\ell m,j),(i+\ell' m,j), (2i+\ell m+\ell' m,2j)\}\colon i\in \mathbf{I},\ 0\leq
j<n,\ 0\leq \ell<\ell'<g\},\\
\vspace{2pt} \D_4=\{\{(0,0),(\ell m,\frac{n}{2}), (i+\ell'n,j),(i+\ell m+\ell'm,\frac{n}{2}+j)\}\colon i\in \mathbf{I},\ 0\leq
j< \frac{n}{2},\ 1\leq \ell\leq \frac{g-1}{2}, 0\leq \ell'<g\}.\\
\end{array}
\]
Let $\D_i'=\{D+\delta:\ D\in \D_i,\ \delta\in \mathbb{Z}_{mg}\times \mathbb{Z}_n\}$ for $1\leq i\leq 4$.
Denote $\D'=\D_1'\cup \D_2'\cup \D_3'\cup \D_4'$. Similar to the proof of Theorem \ref{G*-design}, it is readily checked that $\D'$ is the set of blocks of the required strictly $\mathbb{Z}_{mg}\times \mathbb{Z}_{n}$-invariant $G^*(\frac{m}{e},egn,4,3)$  design relative to $\frac{m}{e}\mathbb{Z}_{mg}\times \mathbb{Z}_n$. \qed

\begin{example}
\label{Z10Z10G(100)}
There is a strictly $\mathbb{Z}_{10}\times \mathbb{Z}_{10}$-invariant $G^*(5,20,4,3)$  design relative to $5\mathbb{Z}_{10}\times \mathbb{Z}_{10}$ and an optimal $(10,10,4,2)$-OOSPC with the size meeting the upper bound $(\ref{2D-JohnsonBound})$.
\end{example}

\proof As it has been pointed out before Construction \ref{G*-design}, the strictly $\mathbb{Z}_{10}\times \mathbb{Z}_{2}$-invariant $G(5,4,4,3)$  design relative to $5\mathbb{Z}_{10}\times \mathbb{Z}_2$ in Example \ref{Z10Z2SQS(20)} is also a strictly $\mathbb{Z}_{10}\times \mathbb{Z}_{2}$-invariant $G^*(5,4,4,3)$  design. Applying Construction \ref{G*-design} with $g=5$ yields a
strictly $\mathbb{Z}_{10}\times \mathbb{Z}_{10}$-invariant $G^*(5,20,4,3)$  design relative to $5\mathbb{Z}_{10}\times \mathbb{Z}_{10}$. Since a strictly $\mathbb{Z}_{10}\times \mathbb{Z}_{2}$-invariant $G^*(5,4,4,3)$  design is also a strictly $\mathbb{Z}_{10}\times \mathbb{Z}_{2}$-invariant $PQS(20)$ with $J(10,2,4,3)$ base blocks, by Construction \ref{filling} there is a strictly $\mathbb{Z}_{10}\times \mathbb{Z}_{10}$-invariant $PQS(100)$ with $J(10,10,4,2)$ base blocks, which leads to an optimal $(10,10,4,2)$-OOSPC with the size meeting the upper bound $(\ref{2D-JohnsonBound})$ by Theorem \ref{equivalence}.
\qed

\section{Constructions of strictly $\mathbb{Z}_m\times \mathbb{Z}_n$-invariant $G(m,n,4,3)$  designs via 1-fan designs}

In this section, use $s$-fan designs admitting an automorphism group to construct strictly $\mathbb{Z}_m\times \mathbb{Z}_n$-invariant $G(m,n,4,3)$  designs.

Let $s$ be a non-negative integer and $K_0$, $K_1,\ldots,K_s$
be sets of positive integers. An {\em $s$-fan design} is an
$(s+3)$-tuple $(X,{\cal G},{\cal B}_0,\ldots,{\cal
B}_s)$ where $X$ is a set of $mn$ points, ${\cal G}$ is a partition of $X$ into
$m$ sets of size $n$ (called {\it groups}) and ${\cal B}_0,\ldots,{\cal
B}_s$ are sets of subsets of $X$ satisfying that each $(X,{\cal G},{\cal B}_i)$ is an $H(m,n,K_i,2)$  design for $0\leq i< s$ and $(X,{\cal G},{\cal B}_0\cup \cdots \cup {\cal B}_s)$ is a $G(m,n,K_1\cup \cdots \cup K_s,3)$  design. For simplicity, it is denoted by
$s$-FG$(3,(K_0,\ldots,K_s),mn)$ of type
$n^m$. Note that a $0$-FG is nothing but a $G$-design. For general $s$-fan designs, the reader is referred to \cite{Hartman1994}.

An {\em automorphism group} of an $s$-fan design $(X,{\cal G},{\cal
B}_0,\ldots, {\cal B}_s)$ is a permutation group
on $X$ leaving ${\cal G}$, ${\cal B}_0, \ldots, {\cal B}_s$ invariant, respectively.
All  automorphisms of an $s$-fan design form a group, called the {\em full automorphism group} of the $s$-fan design.
Any subgroup of the full automorphism group is called an {\em automorphism group} of the $s$-fan design.
An $s$-fan design $(X,\Gamma,{\cal
B}_0,\ldots, {\cal B}_s)$ is said to be $G$-{\em invariant} if it admits $G$ as a point-regular automorphism group. Moreover, it
is said to be {\em strictly $G$-invariant} if it is $G$-invariant and the stabilizer of each $B\in \B$ under $G$ equals the identity of $G$.
For  a  $G$-invariant
$s$-FG, we always identify
the point set $X$ with $G$ and  the automorphisms are
regarded as translations $\sigma_a$ defined by $\sigma_a(x)= x+a$ for $x\in G$, where $a\in G$.
Let $L$ be a subgroup of $G$.
If the group set of a $G$-{\em invariant}  $s$-FG is a set of cosets of $L$ in $G$, then it is a $G$-{\em invariant}  $s$-FG relative to $L$.

\begin{example}
\label{1-FG(3,(3,4),9)}
 $(\mathbb{Z}_3\times \mathbb{Z}_3,\{\{i\}\times \mathbb{Z}_3:i\in \mathbb{Z}_3\},\B_0,\B_1)$ is a $\mathbb{Z}_3\times \mathbb{Z}_3$-invariant $1$-FG$(3,(3,4),9)$ of type $3^3$, where $\B_0$ and $\B_1$ are as follows:
\end{example}
\[
\begin{array}{lllll}
\B_0=\bigcup_{0\leq i\leq 2}\big\{\{(0,i),(1,i),(2,i)\},\{(i,0),(i+1,1)(i+2,2)\},\{(i,0),(i+2,1),(i+1,2)\}\big\},\\
\B_1=\bigcup_{0\leq i,j\leq 2}\big\{\{(i,j+1),(i,j+2),(i+1,j),(i+2,j)\},\{(i,j),(i,j+1),(i+1,j),(i+1,j+1)\}\big\}.\\
\end{array}
\]

An $s$-fan design of type $n^m$ $(X,\Gamma,{\cal
B}_0,\ldots, {\cal B}_s)$ is said to be {\em semi-cyclic} if the $s$-fan design admits an automorphism $\sigma$ consisting of $m$ cycles of length $n$ and leaving each group, ${\cal
B}_0,\ldots, {\cal B}_s$ invariant. For  a semi-cyclic $s$-fan design of type $n^m$,  without loss of generality we can identify
the point set $X$ with $I_m\times \mathbb{Z}_n$, and the automorphism $\sigma$
can be taken as $(i,j)\mapsto (i,j+1)$ $(-,{\rm mod}\ n)$,  $(i,j)\in I_m\times \mathbb{Z}_n$.

A {\em rotational} SQS$(m+1)$ is an SQS$(m+1)$ with an automorphism consisting of a cycle of length $m$ and one fixed point. Such a design is denoted by RoSQS$(m+1)$. As pointed out in \cite{FCJ2009}, there is an equivalence between 1-FGs and RoSQSs as follows.

\begin{lemma}
{\rm \cite{FCJ2009}}
\label{RoSQS-1FG}
An RoSQS$(m+1)$ with $m\equiv 1 \pmod 6$ is equivalent to a strictly cyclic
$1$-FG$(3,(3,4),m)$ of type $1^m$. An RoSQS$(m+1)$ with $m\equiv 3 \pmod6$ is equivalent to a strictly
cyclic $1$-FG$(3,(3,4),m)$ of type $3^{m/3}$.
\end{lemma}

 Bitan and Etzion have pointed out in \cite{BE1993} that the existence
of an RoSQS$(v+1)$ implies the existence of an optimal
$1$-D $(v,4,2)$-OOC. Similarly, we can give the following relationship.

\begin{lemma}
\label{1-FG-OOSPC}
Let $mn\equiv 1,3\pmod 6$. Then there is an optimal $(m,n,4,2)$-OOSPC with the size attaining the upper bound $(\ref{2D-JohnsonBound})$ if and only if there is a $\mathbb{Z}_m\times \mathbb{Z}_n$-invariant $1$-FG$(3,(3,4),mn)$ of type $1^{mn}$.
\end{lemma}

\proof Suppose that $\C$ is an $(m,n,4,2)$-OOSPC with the size attaining the upper bound $(\ref{2D-JohnsonBound})$. By Theorem \ref{equivalence}, there is a strictly $\mathbb{Z}_m\times \mathbb{Z}_n$-invariant PQS$(mn)$ with $\frac{(mn-1)(mn-3)}{24}$ base blocks, whose set of base blocks is denoted by $\B$. Then, there are $\frac{mn(mn-1)}{6}$ triples in the leave ${\cal L}$, and the leave is $\mathbb{Z}_m\times \mathbb{Z}_n$-invariant. Clearly, for any
pair $\{(a_1,b_1),(a_2,b_2)\}$ of $\mathbb{Z}_m\times \mathbb{Z}_n$ there is at least one triple in the leave containing $\{(a_1,b_1),(a_2,b_2)\}$ since $mn-2$ is odd. It follows that there are at least $\frac{mn(mn-1)}{6}$ triples in the leave. Consequently, each pair occurs in exactly one triple in the leave, i.e.,  the leave is the block set of an STS$(mn)$ over $\mathbb{Z}_m\times \mathbb{Z}_n$ admitting $\mathbb{Z}_m\times \mathbb{Z}_n$ as a point-regular automorphism group. So, $(\mathbb{Z}_m\times \mathbb{Z}_n,\{\{x\}:x\in \mathbb{Z}_m\times \mathbb{Z}_n\}, {\cal L},\B)$  is a $\mathbb{Z}_m\times \mathbb{Z}_n$-invariant $1$-FG$(3,(3,4),mn)$ of type $1^{mn}$.

Conversely,  there are $\frac{mn(mn-1)(mn-3)}{24}$ quadruples in a $\mathbb{Z}_m\times \mathbb{Z}_n$-invariant $1$-FG$(3,(3,4),mn)$ of type $1^{mn}$, and all orbits of quadruples are full under $\mathbb{Z}_m\times \mathbb{Z}_n$. Therefore, all quadruples form a strictly $\mathbb{Z}_m\times \mathbb{Z}_n$-invariant PQS$(mn)$, which leads to an $(m,n,4,2)$-OOSPC by Theorem \ref{equivalence} with the size attaining the upper bound $(\ref{2D-JohnsonBound})$. \qed

In \cite{FCJ2009}, Feng et al. showed that there exists a semi-cyclic
$1$-FG$(3,(3,4),3h)$ of type $h^3$ if there exists an RoSQS$(h+1)$. By the necessary condition, $h$ must be odd. It follows that all orbits of quadruples of
a semi-cyclic $1$-FG$(3,(3,4),3h)$ of type $h^3$ are full. Since the blocks of size three are from three distinct groups, all block-orbits of size three in a semi-cyclic $1$-FG$(3,(3,4),3h)$ are also full. So, a semi-cyclic
$1$-FG$(3,(3,4),3h)$ of type $h^3$ must be strictly semi-cyclic.

\begin{lemma}
\label{strictly semi-cyclic 1-FG(3,(3,4),3h)}
If there exists an RoSQS$(h+1)$, then there exists a strictly semi-cyclic
$1$-FG$(3,(3,4),3h)$ of type $h^3$.
\end{lemma}

Hartman established a fundamental construction for 3-designs \cite{Hartman1994}.  By using it, Hartman gave a new existence proof of Steiner quadruple systems.
The following is a special case.

\begin{theorem}
{\rm \cite{Hartman1994}}
\label{FC-Hartman}
Suppose there is a $1$-FG$(3,(K_0,K_1),mn)$ of type $n^m$
$($called a master design$)$. If there exists an
 $s$-FG$(3,(L_0,L_1,\ldots,L_s),gk)$ of type
$g^{k}$ for any $k\in K_0$ and an $H(k,g,L_s,3)$  design for
any $k\in K_s$, then there exists an
$s$-FG$(3,(L_0,\ldots,L_s),mng)$ of type $(ng)^m$.
\end{theorem}

By using Theorem \ref{FC-Hartman}, Feng et al. established a recursive construction for strictly cyclic $s$-fan designs \cite{FCJ2008}. We generalize it as follows. The detailed proof Construction \ref{FC-sFG} is moved to Appendix B.

\begin{construction}
\label{FC-sFG} Suppose there
is a strictly $\mathbb{Z}_{m}\times \mathbb{Z}_n$-invariant $1$-FG$(3,(K_0,K_1),mn)$ of type $(en)^{m/e}$
relative to $\frac{m}{e}\mathbb{Z}_{m}\times \mathbb{Z}_n$ $($called a master design$)$. If there exists a
strictly semi-cyclic $s$-FG$(3,(L_0,L_1,\ldots,L_s),gk)$ of type
$g^{k}$ for any $k\in K_0$, and a semi-cyclic $H(k,g,L_s,3)$  design for any
$k\in K_1$, then there exists a strictly $\mathbb{Z}_{m}\times \mathbb{Z}_{ng}$-invariant
$s$-FG$(3,(L_0,\ldots,L_s),mng)$ of type $(eng)^{m/e}$ relative to $\frac{m}{e}\mathbb{Z}_{m}\times \mathbb{Z}_{ng}$ and a strictly $\mathbb{Z}_{mg}\times \mathbb{Z}_{n}$-invariant
$s$-FG$(3,(L_0,\ldots,L_s),mng)$ of type $(eng)^{m/e}$ relative to $\frac{m}{e}\mathbb{Z}_{mg}\times \mathbb{Z}_{n}$.
\end{construction}

\proof Let $(\mathbb{Z}_{m}\times \mathbb{Z}_n,{\cal G},{\cal B}_0,{\cal B}_1)$ be a strictly $\mathbb{Z}_{m}\times \mathbb{Z}_n$-invariant
$1$-FG$(3,(K_0,K_1),mn)$ of type $(en)^{m/e}$ where ${\cal
G}=\{\{i,i+\frac{m}{e},\ldots,i+m-\frac{m}{e}\}\times \mathbb{Z}_n : 0\leq i<\frac{m}{e}\}$.
Denote the family of base blocks of this design by ${\cal F}={\cal
F}_0\bigcup {\cal F}_1$, where ${\cal F}_0$ and ${\cal F}_1$
generate all blocks of ${\cal B}_0$ and ${\cal B}_1$, respectively.

For each base block $B\in {\cal F}_0$, construct a strictly
semi-cyclic $s$-FG$(3,(L_0,\ldots,L_s),|B|g)$ of type
$g^{|B|}$ on $B\times \mathbb{Z}_g$ with group set $\{\{x\}\times \mathbb{Z}_g: x\in
B\}$. Denote the family of base blocks of the $j$-th subdesign
$H(|B|,g,L_j,2)$  design by ${\cal A}_B^j$ for $0\leq j< s$, and denote
the family of all the other base blocks by ${\cal A}_B^s$. Let ${\cal
A}_B=\bigcup_{j=0}^s {\cal A}_B^j$.

For each base block $B\in {\cal F}_1$, construct a semi-cyclic
$H(|B|,g,L_s,3)$  design on $B\times \mathbb{Z}_g$ with groups $\{\{x\}\times \mathbb{Z}_g:
x\in B\}$. Denote the family of base blocks of this design by
${{\cal D}_B}$.

Let ${\cal A}_j=\bigcup_{B\in
{\cal F}_0} {\cal A}_B^j$ for $0\leq j <s$ and ${\cal A}_s =
(\bigcup_{B\in
{\cal F}_0} {\cal A}_B^s) \bigcup (\bigcup_{B\in {\cal
F}_1} {{\cal D}_B})$, and ${\cal G}'=\{\{i,i+\frac{m}{e},\ldots,i+m-\frac{m}{e}\}\times \mathbb{Z}_{ng}\colon 0\leq i<\frac{m}{e}\}$.
Define a mapping $\tau$ from $\mathbb{Z}_{m}\times \mathbb{Z}_{n}\times \mathbb{Z}_g$ to $\mathbb{Z}_{m}\times \mathbb{Z}_{ng}$ by
$\tau(x,y,z)=(x,y+zn).$
Now we construct a strictly $\mathbb{Z}_{m}\times \mathbb{Z}_{ng}$-invariant
$s$-FG$(3,(L_0,\ldots,L_s),mng)$ of type $(eng)^{m/e}$ as
follows: For each $C\in(\bigcup_{0\leq j\leq s}{\cal A}_j)$, define $\tau(C)=\{\tau(c):c\in C\}$. For $0\leq j \leq s$, let
$$\hspace{1cm}  {{\cal A}_j^*}=\bigcup_{_{C\in {\cal A}_j}} \tau(C),$$
$$
\hspace{1cm} {{\cal A}'_j}=\{A+\delta:A\in {\cal A}_j^*,\delta\in
\mathbb{Z}_{m}\times \mathbb{Z}_{ng}\},$$ where $A+\delta=\{u+\delta: u\in A\}$.
We claim that $(\mathbb{Z}_{m}\times \mathbb{Z}_{ng},{\cal G}',{{\cal A}'_0},\ldots,{{\cal
A}'_s})$ is a strictly $\mathbb{Z}_{m}\times \mathbb{Z}_{ng}$-invariant
$s$-FG$(3,(L_0,\ldots,L_s),mng)$ of type $(eng)^{m/e}$.

Let ${\cal G}^{''}=\{\{i,i+\frac{m}{e},\ldots,i+mg-\frac{m}{e}\}\times \mathbb{Z}_{n}\colon 0\leq i<\frac{m}{e}\}$.
Define a mapping $\varphi$ from $\mathbb{Z}_{m}\times \mathbb{Z}_{n}\times \mathbb{Z}_g$ to $\mathbb{Z}_{mg}\times \mathbb{Z}_{n}$ by
$\tau(x,y,z)=(x+zm,y).$
Now we construct a strictly $\mathbb{Z}_{mg}\times \mathbb{Z}_{n}$-invariant
$s$-FG$(3,(L_0,\ldots,L_s),mng)$ of type $(eng)^{m/e}$ as
follows: For each $C\in(\bigcup_{0\leq j\leq s}{\cal A}_j)$, define $\varphi(C)=\{\varphi(c):c\in C\}$. For $0\leq j \leq s$, let
$$\hspace{1cm}  {\cal A}_j^{**}=\bigcup_{_{C\in {\cal A}_j}} \varphi(C),$$
$$
\hspace{1cm} {{\cal A}^{''}_j}=\{A+\delta:A\in {\cal A}_j^{**},\delta\in
\mathbb{Z}_{mg}\times \mathbb{Z}_{n}\},$$ where $A+\delta=\{u+\delta: u\in A\}$.
Similarly, it is readily checked that $(\mathbb{Z}_{mg}\times \mathbb{Z}_{n},{\cal G}^{''},{{\cal A}^{''}_0},\ldots,{{\cal
A}^{''}_s})$ is a strictly $\mathbb{Z}_{mg}\times \mathbb{Z}_{n}$-invariant
$s$-FG$(3,(L_0,\ldots,L_s),mng)$ of type $(eng)^{m/e}$. \qed

\begin{corollary}
\label{G+semi-cyclic G(2,g,4,3)}
Suppose that  there is a strictly $\mathbb{Z}_{m}\times \mathbb{Z}_n$-invariant $G(\frac{m}{e},en,4,3)$  design relative to $\frac{m}{e}\mathbb{Z}_{m}\times \mathbb{Z}_n$ such that all elements of order $2$ are contained in $\frac{m}{e}\mathbb{Z}_m\times \mathbb{Z}_n$.  If there is a
strictly semi-cyclic $G(2,g,4,3)$  design, then there exists a strictly $\mathbb{Z}_{m}\times \mathbb{Z}_{ng}$-invariant
$G(\frac{m}{e},emg,4,3)$  design relative to $\frac{m}{e}\mathbb{Z}_{m}\times \mathbb{Z}_{ng}$ and a  strictly $\mathbb{Z}_{mg}\times \mathbb{Z}_{n}$-invariant
$G(\frac{m}{e},emg,4,3)$  design relative to $\frac{m}{e}\mathbb{Z}_{mg}\times \mathbb{Z}_n$.
\end{corollary}

\proof Let $\B_1$ be the set of blocks of a strictly $\mathbb{Z}_{m}\times \mathbb{Z}_n$-invariant $G(\frac{m}{e},en,4,3)$  design on $\mathbb{Z}_{m}\times \mathbb{Z}_n$ with group set ${\cal
G}=\{\{i,i+\frac{m}{e},\ldots,i+m-\frac{m}{e}\}\times \mathbb{Z}_n : 0\leq i<\frac{m}{e}\}$.
Let $\F_0=\{\{(0,0),(i,j)\}:1\leq i\leq \lfloor\frac{m}{2}\rfloor, i\not\equiv 0\pmod {\frac{m}{e}},j\in \mathbb{Z}_n\}$ and $\B_0=\{P+\delta:P\in \F_0, \delta\in \mathbb{Z}_m\times \mathbb{Z}_n\}$. Since all elements of order $2$ are contained in $\frac{m}{e}\mathbb{Z}_m\times \mathbb{Z}_n$, the quadruple $(X,\G,\B_0,\B_1)$ is a strictly  $\mathbb{Z}_{m}\times \mathbb{Z}_n$-invariant  $1$-FG$(3,(2,4),mn)$ of type $(en)^{m/e}$. Since there is a
strictly semi-cyclic $G(2,g,4,3)$  design by assumption and a semi-cyclic $H(4,g,4,3)$  design by Lemma \ref{semi-cyclic H(6n+2,m,4,3)}, applying Construction \ref{FC-sFG} yields a strictly $\mathbb{Z}_{m}\times \mathbb{Z}_{ng}$-invariant
$G(\frac{m}{e},emg,4,3)$  design and a  strictly $\mathbb{Z}_{mg}\times \mathbb{Z}_{n}$-invariant
$G(\frac{m}{e},emg,4,3)$  design. \qed

\begin{corollary}
\label{RoSQS-OOSPC} Suppose there is an RoSQS$(m+1)$ and an RoSQS$(n+1)$. If $m\equiv 1\pmod 6$ then there is a strictly $\mathbb{Z}_m\times \mathbb{Z}_n$-invariant $1$-FG$(3,(3,4),mn)$ of type $n^{m}$ relative to $\{0\}\times \mathbb{Z}_n$. If $m\equiv 3\pmod 6$ then there is a strictly $\mathbb{Z}_m\times \mathbb{Z}_n$-invariant $1$-FG$(3,(3,4),mn)$ of type $(3n)^{m/3}$ relative to $\frac{m}{3}\mathbb{Z}_{m}\times \mathbb{Z}_n$.
\end{corollary}

\proof Since there is an RoSQS$(m+1)$ by assumption, there is a strictly cyclic $1$-FG$(3,(3,4),m)$ of type $1^m$ if $m\equiv 1\pmod 6$, a strictly cyclic $1$-FG$(3,(3,4),m)$ of type $3^{m/3}$ if $m\equiv 3\pmod 6$ by Lemma \ref{RoSQS-1FG}. Since there is an RoSQS$(n+1)$, there is a strictly semi-cyclic $1$-$FG(3,(3,4),3n)$ of type $n^3$ by Lemma \ref{strictly semi-cyclic 1-FG(3,(3,4),3h)}. Also, there is  a
semi-cyclic $H(4,n,4,3)$  design by Lemma \ref{semi-cyclic H(6n+2,m,4,3)}. Therefore,  applying Construction \ref{FC-sFG} gives a strictly $\mathbb{Z}_m\times \mathbb{Z}_n$-invariant $1$-FG$(3,(3,4),mn)$ of type $n^{m}$ if  $m\equiv 1\pmod 6$,  a strictly $\mathbb{Z}_m\times \mathbb{Z}_n$-invariant $1$-FG$(3,(3,4),mn)$ of type $(3n)^{m/3}$
if $m\equiv 3\pmod 6$.
\qed

\begin{corollary}
\label{RoSQS-ZmZn-G-desings}
If there is an RoSQS$(m+1)$ and a
strictly semi-cyclic $G(3,g,4,3)$  design, then there exists a strictly $\mathbb{Z}_{m}\times \mathbb{Z}_{g}$-invariant
$G(m,g,4,3)$  design relative to $\{0\}\times \mathbb{Z}_g$ if $m\equiv 1\pmod 6$, and a strictly $\mathbb{Z}_{m}\times \mathbb{Z}_{g}$-invariant
$G(\frac{m}{3},3g,4,3)$  design relative to $\frac{m}{3}\mathbb{Z}_{m}\times \mathbb{Z}_g$ if $m\equiv 3\pmod 6$.
\end{corollary}

\proof Since there is an RoSQS$(m+1)$ by assumption, there is a strictly cyclic $1$-FG$(3,(3,4),m)$ of type $1^m$ if $m\equiv 1\pmod 6$, a strictly cyclic $1$-FG$(3,(3,4),m)$ of type $3^{m/3}$ if $m\equiv 3\pmod 6$ by Lemma \ref{RoSQS-1FG}. Since there is a strictly semi-cyclic $G(3,g,4,3)$  design by assumption and  a
semi-cyclic $H(4,g,4,3)$  design by Lemma \ref{semi-cyclic H(6n+2,m,4,3)},  applying Construction \ref{FC-sFG} gives the conclusion.   \qed

\section{New $(m,n,4,2)$-OOSPCs}
In this section, we use constructions in Sections 3, 4 and 5 to establish new optimal $(m,n,4,2)$-OOSPCs.

Since the survey of Lindner and Rosa \cite{LR1978},  many recursive constructions for cyclic SQSs have been given, including the doubling construction,   product constructions. Recently, Feng et al. established some recursive constructions for strictly
cyclic 3-designs, as corollaries, many known constructions for strictly cyclic Steiner quadruple
systems were unified \cite{FCJ2008}. The work of K\"{o}hler on $S$-cyclic SQS has been extended by Bitan  and Etzion \cite{BE1993}, Siemon \cite{siemon1987}, \cite{siemon1989}, \cite{siemon1991}, \cite{siemon1998}. Although a great deal has been done on cyclic SQSs, the spectrum remains wide open.

A cyclic SQS$(v)$ $(\mathbb{Z}_v,\B)$ is said to be $S$-cyclic if each block satisfies $-B=B+a$ for some $a\in \mathbb{Z}_v$. Piotrowski gave necessary and sufficient conditions for the existence of an $S$-cyclic SQS$(v)$ \cite{Piotrowski1985}.
\begin{theorem}
\label{S-SQS(v)}
\emph{\cite{Piotrowski1985}}
\label{S-CSQS}
An $S$-cyclic $SQS(v)$ exists if and only if $v\equiv 0 \pmod{2},  v\not\equiv 0 \pmod{3},  v\not\equiv 0 \pmod{8},  v\ge 4$
 and if for any prime divisor $p$ of $v$ there exists an $S$-cyclic $SQS(2p)$.
\end{theorem}

\begin{theorem}
\emph{\cite{BE1993}}
\label{S-CSQS(4p)}
For any prime $p\equiv 5\pmod{12}$ with $p<1500000$, there is an  $S$-cyclic $SQS(4p)$.
\end{theorem}

\begin{theorem}
\emph{\cite{LJ-appear}}
\label{small-S-CSQS}
For each prime $p\equiv 1\pmod 4$ with $p\leq 10^5$, there is an $S$-cyclic $SQS(2p)$.
\end{theorem}

Let $n$ be even and $mn\equiv n\pmod 4$. In a cyclic $G(m,n,4,3)$  design, if any triple of form $\{j,j+i,j+2i\}$ or
$\{j,j+i,j+mn/2\}$, where $1\leq i\leq mn/2$, $i\not\equiv 0\pmod
m$ and $0\leq j\leq mn-1$, is contained in the block
$\{j,j+a,j-a,j+\frac{mn}{2}\}$ for some $1\leq a\leq
\lfloor\frac{mn}{4}\rfloor$ and $a\not\equiv 0\pmod m$, then such
a cyclic $G$-design is denoted by cyclic $G^*(m,n,4,3)$  design. As pointed out in \cite{FCJ2008}, a cyclic
$G^*(m,n,4,3)$ is always strictly cyclic. The following recursive construction for cyclic $G^*$-designs was given in \cite{FCJ2008}.
For the completeness, we describe how to construct a cyclic $G^*(m,ng,4,3)$  design from a cyclic $G^*(m,n,4,3)$  design here.

\begin{theorem}
\emph{\cite{FCJ2008}}
\label{Product-Cyclic-G-design}  If there
exists a cyclic $G^*(m,n,4,3)$  design, then
there exists a cyclic $G^*(m,ng,4,3)$  design for any odd integer $g$.
\end{theorem}

\proof Let $(\mathbb{Z}_{mn},\{\{i,i+m,\ldots,i+mm-m\}:0\leq i<m\},{\cal B})$ be a cyclic $G^*(m,n,4,3)$  design. Denote the
family of base blocks of this design by ${\cal F}={\cal F}_1\bigcup
{\cal F}_2$, where ${\cal F}_1$ consists of all base blocks in the
form of $\{0,\frac{mn}{2},i,-i,\}$, $1\leq i\leq \lfloor\frac{mn}{4}\rfloor$ and $i\not\equiv 0\pmod m$, and ${\cal
F}_2$ consists of all the other base blocks. It is easy to see that
$|\F_1|=n(m-1)/4$ and $|\F_2|=n(m-1)(mn+n-9)/24$.

Define
\[
\begin{array}{l}
\vspace{2pt} \D_1=\{\{i_0,i_1+k_1mn,i_2+k_2mn,i_3+k_1mn+k_2mn\}:\{i_0,i_1,i_2,i_3\}\in \F_2,~0\le k_1,k_2<g\},\\
\vspace{2pt} \D_2=\{\{0,i,-i,\frac{mng}{2}\}\colon 1\leq i\leq \frac{mng}{4}, i\not\equiv 0\pmod m\},\\
\vspace{2pt} \D_3=\{\{0,i+\ell mn,i+\ell' mn,2i+\ell mn+\ell' mn\}\colon 1\leq i\leq \frac{mn}{2}, i\not\equiv 0\pmod m,\ 0\leq \ell<\ell'<g\},\\
\vspace{2pt} \D_4=\{\{0,\frac{mn}{2}+\ell mn,i+\ell'mn, i+\frac{mn}{2}+\ell mn+\ell'mn\}\colon 1\leq i\leq \frac{mn}{4}, i\not\equiv 0\pmod m,\\
\hspace{8.6cm} 0\leq \ell<\frac{g-1}{2}, 0\leq \ell'<g\}.\\
\end{array}
\]
Let $\D_i'=\{D+\delta:\ D\in \D_i,\ \delta\in \mathbb{Z}_{mng}\}$ for $1\leq i\leq 4$ and
 $\D'=\D_1'\cup \D_2'\cup \D_3'\cup \D_4'$. Then $\D'$ is the set of blocks of the required cyclic  $G^*(m,ng,4,3)$  design on $\mathbb{Z}_{mng}$ with
the group set $\{\{i,i+m,\ldots,i+mng-m\}:\ 0\leq i<m\}$. \qed

\begin{theorem}
\label{S-ZmZnSQS(mn)-OOSPC}
Let $m,n,g$ be odd integers such that there is an $S$-cyclic SQS$(2p)$ for each prime divisor $p$ of $m$ and $n$. If there is an optimal  $1$-D $(2^{\epsilon}g,4,2)$-OOC with $J(1,2^{\epsilon}g,4,2)$ codewords, then there is an optimal $(m,2^{\epsilon}ng,4,2)$-OOSPC and an optimal $(mg,2^{\epsilon}n,4,2)$-OOSPC with the size attaining the upper bound $(\ref{2D-JohnsonBound})$, where $\epsilon\in \{1,2\}$.
\end{theorem}

\proof  Since there is an $S$-cyclic SQS$(2p)$ for each prime divisor $p$ of $m$ and $n$, there is an $S$-cyclic SQS$(2^{\varepsilon}m)$ and an $S$-cyclic SQS$(2^{\varepsilon}n)$ by Theorem \ref{S-SQS(v)}, where $\epsilon\in \{1,2\}$. Since $\mathbb{Z}_{2^{\varepsilon}m}$ is isomorphic to $\mathbb{Z}_m\times \mathbb{Z}_{2^{\epsilon}}$, the existence of an $S$-cyclic SQS$(2^{\varepsilon}m)$ implies that there is a strictly $\mathbb{Z}_m\times \mathbb{Z}_{2^{\epsilon}}$-invariant $G^*(m,2^{\varepsilon},4,3)$  design. Applying Construction \ref{G*-design} gives a strictly $\mathbb{Z}_m\times \mathbb{Z}_{2^{\epsilon}ng}$-invariant $G^*(m,2^{\varepsilon}ng,4,3)$  design relative to $\{0\}\times \mathbb{Z}_{2^{\epsilon}ng}$ and  applying Construction \ref{G*-design-1} gives a strictly $\mathbb{Z}_{mg}\times \mathbb{Z}_{2^{\epsilon}n}$-invariant $G^*(m,2^{\varepsilon}ng,4,3)$  design relative to $m\mathbb{Z}_{mg}\times \mathbb{Z}_{2^{\epsilon}n}$.

Since an $S$-cyclic SQS$(2^{\varepsilon}n)$ implies the existence of a cyclic $G^*(n,2^{\epsilon},4,3)$  design, applying Theorem \ref{Product-Cyclic-G-design}  gives a cyclic $G^*(n,2^{\varepsilon}g,4,3)$  design. Since there is an optimal $1$-D $(2^{\epsilon}g,4,2)$-OOC with $J(1,2^{\epsilon}g,4,2)$ codewords by assumption which corresponds to  a strictly cyclic PQS$(2^{\epsilon}g)$ with $J(1,2^{\epsilon}g,4,2)$ base blocks, there is a strictly cyclic PQS$(2^{\epsilon}ng)$ with $J(1,2^{\varepsilon}ng,4,2)$ base blocks by Construction \ref{filling}. When we input this cyclic PQS$(2^{\epsilon}ng)$ into the strictly $\mathbb{Z}_m\times \mathbb{Z}_{2^{\epsilon}ng}$-invariant $G^*(m,2^{\varepsilon}ng,4,3)$  design,  applying Construction \ref{filling} gives a strictly  $\mathbb{Z}_m\times \mathbb{Z}_{2^{\epsilon}ng}$-invariant PQS$(2^{\epsilon}mng)$ with $J(m,2^{\epsilon}ng,4,2)$ base  blocks, which leads to an optimal $(m,2^{\varepsilon}ng,4,2)$-OOSPC with the size attaining the upper bound $(\ref{2D-JohnsonBound})$.

Since an $S$-cyclic SQS$(2^{\varepsilon}n)$ implies the existence of a strictly $\mathbb{Z}_{n}\times \mathbb{Z}_{2^{\epsilon}}$-invariant $G^*(n,2^{\varepsilon},4,3)$  design, applying Construction \ref{G*-design} gives a strictly $\mathbb{Z}_{n}\times \mathbb{Z}_{2^{\epsilon}g}$-invariant $G^*(n,2^{\varepsilon}g,4,3)$  design.
Since there is a strictly cyclic PQS$(2^{\epsilon}g)$ with $J(1,2^{\epsilon}g,4,2)$ base blocks, there is a strictly  $\mathbb{Z}_{n}\times \mathbb{Z}_{2^{\epsilon}g}$-invariant PQS$(2^{\epsilon}ng)$ with $J(n,2^{\varepsilon}g,4,2)$ base blocks by Construction \ref{filling}. Since $\mathbb{Z}_{n}\times \mathbb{Z}_{2^{\epsilon}g}$ is isomorphic to $\mathbb{Z}_{2^{\epsilon}n}\times \mathbb{Z}_{g}$, there is a strictly $\mathbb{Z}_{2^{\epsilon}n}\times \mathbb{Z}_{g}$-invariant PQS$(2^{\epsilon}ng)$ with $J(2^{\varepsilon}n,g,4,2)$ base blocks. Further, we put this PQS into the strictly $\mathbb{Z}_{mg}\times \mathbb{Z}_{2^{\epsilon}n}$-invariant $G^*(m,2^{\varepsilon}ng,4,3)$  design relative to $m\mathbb{Z}_{mg}\times \mathbb{Z}_{2^{\epsilon}n}$. By applying Construction \ref{filling} we obtain a strictly  $\mathbb{Z}_{mg}\times \mathbb{Z}_{2^{\epsilon}n}$-invariant PQS$(2^{\epsilon}mng)$ with $J(mg,2^{\epsilon}n,4,2)$ base  blocks, which leads to an optimal $(mg,2^{\varepsilon}n,4,2)$-OOSPC with the size attaining the upper bound $(\ref{2D-JohnsonBound})$. \qed

\begin{lemma}
\emph{\cite{CC2004,FCJ2008}}
\label{Small-OOC}
There is an optimal $1$-D $(n,4,2)$-OOC with $J(1,n,4,2)$ codewords for all $7\leq n\leq 100$ with the definite exceptions of $n \in \{9$, $12$,
$13$, $24$, $48$, $72$, $96\}$ and possible exceptions of $n\in
\{45$, $47$, $53$, $55$, $59$, $60$, $65$, $66$, $69$, $71$, $76$,
$77$, $81$, $83$, $84$, $85$, $89$, $91$, $92$, $95$, $97$, $99\}$. There is an optimal $1$-D $(n,4,2)$-OOC with $J(1,n,4,2)-1$ codewords for each $n \in \{9$, $12$,
$13$, $24$, $48$, $72$, $96\}$.
\end{lemma}

\begin{corollary}
\label{(m,2n,4,2)}
Let  $m$ and $n$ be composite numbers whose prime divisors each belong to $\{p\equiv 1\pmod{12}: p\ {\rm is\ a\ prime},\ p<10^5\}\cup \{p\equiv 5\pmod{12}: p\ {\rm is\ a\ prime},\ p<1500000\}$. Then, there is an optimal $(m,2ng,4,2)$-OOSPC $($resp. $(mg,2n,4,2)$-OOSPC$)$ with the size attaining the upper bound $(\ref{2D-JohnsonBound})$ for $g\in \{1,3,5,\ldots$, $49\}\setminus \{33\}$, and an optimal $(m,4ng,4,2)$-OOSPC $($resp. $(mg,4n,4,2)$-OOSPC$)$ with the size attaining the upper bound $(\ref{2D-JohnsonBound})$ for $g\in \{1,3,5,\ldots,13\}\setminus \{3\}$.
\end{corollary}

\proof  Since there is an  $S$-cyclic SQS$(2p)$ for any prime divisor $p$ of $m$ and $n$ by Theorems \ref{S-CSQS(4p)}-\ref{small-S-CSQS} and an optimal
 $1$-D $(2g,4,2)$-OOC with $J(1,2g,4,2)$ codewords for $g\in \{1,3,5,\ldots,49\}\setminus \{33\}$ by Lemma \ref{Small-OOC}, applying Theorem \ref{S-ZmZnSQS(mn)-OOSPC} gives
 an optimal $(m,2ng,4,2)$-OOSPC (resp. $(mg,2n,4,2)$-OOSPC) with the size attaining the upper bound $(\ref{2D-JohnsonBound})$. Similarly, since there is  an optimal $1$-D $(4g,4,2)$-OOC with $J(1,4g,4,2)$ codewords for $g\in \{1,3,5,\ldots,13\}\setminus \{3\}$ by Lemma \ref{Small-OOC}, applying Theorem \ref{S-ZmZnSQS(mn)-OOSPC} gives
 an optimal $(m,4ng,4,2)$-OOSPC (resp. $(mg,4n,4,2)$-OOSPC)  with the size attaining the upper bound $(\ref{2D-JohnsonBound})$. \qed

{\bf Remark}: The optimal $(m,2^{\epsilon}n,4,2)$-OOSPC in \cite{Sawa2010} is obtained again in Corollary \ref{(m,2n,4,2)}.   Comparing with Sawa's method, our construction seems easier.

\begin{lemma}
\label{Z3Z12-PQS(36)}
There exists an optimal $(3,12,4,2)$-OOSPC with the size attaining the upper bound in Lemma $\ref{Bound72k+18,36}$.
\end{lemma}

\proof The following 48 base blocks generate the block set of a strictly $\mathbb{Z}_3\times \mathbb{Z}_{12}$-invariant PQS$(36)$, which corresponds to an optimal $(3,12,4,2)$-OOSPC with the size attaining the upper bound in Lemma $\ref{Bound72k+18,36}$.

\[
\begin{array}{ll}
\{(0,0),(0,1),(0,11),(0,6)\}, & \{(0,0),(1,1),(2,11),(0,6)\}, \\
\{(0,0),(1,3),(2,9),(0,6)\}, & \{(0,0),(1,5),(2,7),(0,6)\}, \\
\{(0,0),(0,1),(0,3),(0,4)\}, & \{(0,0),(0,1),(0,5),(0,8)\}, \\
\{(0,0),(0,1),(1,0),(1,1)\}, & \{(0,0),(0,1),(1,2),(1,3)\}, \\
\{(0,0),(0,1),(1,4),(1,5)\}, & \{(0,0),(0,1),(1,6),(1,7)\}, \\
\{(0,0),(0,1),(1,8),(1,9)\}, & \{(0,0),(0,1),(1,10),(1,11)\}, \\
\{(0,0),(0,2),(0,5),(1,0)\}, & \{(0,0),(0,10),(0,7),(2,0)\}, \\
\{(0,0),(0,2),(0,6),(1,2)\}, & \{(0,0),(0,10),(0,6),(2,10)\}, \\
\{(0,0),(0,2),(1,1),(1,3)\}, & \{(0,0),(0,2),(1,4),(1,6)\}, \\
\{(0,0),(0,2),(1,5),(1,7)\}, & \{(0,0),(0,2),(1,8),(1,10)\}, \\
\{(0,0),(0,2),(1,9),(1,11)\}, & \{(0,0),(0,3),(1,0),(1,3)\}, \\
\{(0,0),(0,3),(1,1),(1,4)\}, & \{(0,0),(0,3),(1,2),(1,5)\}, \\
\{(0,0),(0,3),(1,6),(1,9)\}, & \{(0,0),(0,3),(1,7),(2,4)\}, \\
\{(0,0),(0,9),(2,5),(1,8)\}, & \{(0,0),(0,3),(1,8),(2,7)\}, \\
\{(0,0),(0,4),(1,1),(1,5)\}, & \{(0,0),(0,4),(1,2),(1,10)\}, \\
\{(0,0),(0,4),(1,3),(1,8)\}, & \{(0,0),(0,8),(2,9),(2,4)\}, \\
\{(0,0),(0,4),(1,4),(2,5)\}, & \{(0,0),(0,8),(2,8),(1,7)\}, \\
\{(0,0),(0,4),(1,6),(2,10)\}, & \{(0,0),(0,4),(1,7),(2,9)\}, \\
\{(0,0),(0,4),(1,9),(2,7)\}, & \{(0,0),(0,5),(1,1),(2,4)\}, \\
\{(0,0),(0,5),(1,2),(2,3)\}, & \{(0,0),(0,5),(1,3),(2,2)\}, \\
\{(0,0),(0,5),(1,5),(2,8)\}, & \{(0,0),(0,7),(2,7),(1,4)\}, \\
\{(0,0),(0,5),(1,6),(2,11)\}, & \{(0,0),(0,5),(1,7),(2,7)\}, \\
\{(0,0),(0,7),(2,5),(1,5)\}, & \{(0,0),(0,5),(1,11),(2,6)\}, \\
\{(0,0),(0,6),(1,0),(2,8)\}, & \{(0,0),(0,6),(2,0),(1,4)\}. \\
\end{array}
\]
\qed

\begin{theorem}
\label{(3m,bn,4,2)}
Let $m,n$ be equal to $1$ or the composite numbers of primes as in Corollary $\ref{(m,2n,4,2)}$. Then there is an optimal $(3m,bn,4,2)$-OOSPC with $J(3m,bn,4,2)-1$ codewords attaining the upper bound in Lemma $\ref{Bound72k+18,36}$  for $b\in \{6,12\}$.
\end{theorem}

\proof Start with a strictly $\mathbb{Z}_m\times \mathbb{Z}_{2^{\epsilon}\cdot3n}$-invariant $G^*(m,2^{\varepsilon}\cdot3n,4,3)$  design relative to $\{0\}\times \mathbb{Z}_{2^{\epsilon}\cdot3n}$, $\epsilon\in \{1,2\}$, which exists from the proof of Theorem \ref{S-ZmZnSQS(mn)-OOSPC}. Applying Construction \ref{G*-design-1} gives a strictly $\mathbb{Z}_{3m}\times \mathbb{Z}_{2^{\epsilon}\cdot3n}$-invariant $G^*(m,2^{\varepsilon}\cdot9n,4,3)$  design relative to $m\mathbb{Z}_{3m}\times \mathbb{Z}_{2^{\epsilon}\cdot3n}$. Similarly, there is a strictly $\mathbb{Z}_{3n}\times \mathbb{Z}_{2^{\epsilon}\cdot3}$-invariant $G^*(n,2^{\epsilon}\cdot9,4,3)$  design relative to $n\mathbb{Z}_{3n}\times \mathbb{Z}_{2^{\epsilon}\cdot3}$. Since $\mathbb{Z}_{3n}\times \mathbb{Z}_{2^{\epsilon}\cdot3}$ is isomorphism to $\mathbb{Z}_{2^{\varepsilon}\cdot3n}\times \mathbb{Z}_3$, there is a strictly $\mathbb{Z}_{2^{\epsilon}\cdot3n}\times \mathbb{Z}_3$-invariant $G^*(n, 2^{\epsilon}\cdot9,4,3)$  design relative to $n\mathbb{Z}_{2^{\epsilon}\cdot3n}\times \mathbb{Z}_3$. Since there is an optimal $(3,6,4,2)$-OOSPC  from \cite{Sawa2010} and an optimal $(3,12,4,2)$-OOSPC by  Lemma \ref{Z3Z12-PQS(36)} with the size attaining the upper bound in Lemma $\ref{Bound72k+18,36}$  which is equivalent to a strictly $\mathbb{Z}_3\times \mathbb{Z}_b$-invariant PQS$(3b)$ for $b\in \{6,12\}$, applying Construction \ref{filling} yields a strictly $\mathbb{Z}_{3}\times \mathbb{Z}_{bn}$ PQS$(3bn)$ with the size attaining the upper bound in Lemma $\ref{Bound72k+18,36}$. Further, input this PQS into the strictly $\mathbb{Z}_{3m}\times \mathbb{Z}_{2^{\epsilon}\cdot3n}$-invariant $G^*(m,2^{\epsilon}\cdot9n,4,3)$  design and apply Construction \ref{filling}.  We then obtain an optimal
strictly $\mathbb{Z}_{3m}\times \mathbb{Z}_{bn}$-invariant PQS$(3bmn)$ with the size attaining the upper bound in Lemma $\ref{Bound72k+18,36}$, which leads to an optimal $(3m,3\cdot 2^{\epsilon}n,4,2)$-OOSPC.  \qed

\begin{lemma}
{\rm \cite{FCJ2008}}
\label{Semi-Cyclic-G(2,g,4,3)}
There exists a strictly cyclic $G(2,g,4,3)$  design for each integer $g\equiv 0 \pmod 8$.
\end{lemma}

\begin{lemma}
\emph{\cite[Corollary 6.21]{FCJ2008}}
\label{(24n,4,2)-OOC}
Suppose there is a cyclic SQS$(n)$ with $n\equiv 2$ or $10\pmod {12}$ and a strictly cyclic PQS$(g)$ of the size $J(1,g,4,2)-1$ attaining the upper bound in Lemma $\ref{bound24k}$ with $g\equiv 0\pmod {24}$. Then there is a strictly cyclic PQS$(2^a3^b5^cn^dg)$ of the size $J(1,2^a3^b5^cn^dg,4,2)-1$ attaining the upper bound in Lemma $\ref{bound24k}$ for any nonnegative integers $a,b,c,d$.
\end{lemma}

\begin{theorem}
\label{(m,2a3bn,4,2)}
Let $m,n$ be equal to $1$ or the composite numbers of primes as in Corollary $\ref{(m,2n,4,2)}$. Then there is an optimal $(m,2^a3^bn,4,2)$-OOSPC with $J(m,2^a3^bn,4,2)-1$ codewords attaining the upper bound in Lemma $\ref{bound24k}$  for $a\geq 4, b\geq 1$.
\end{theorem}

\proof Start with  a strictly $\mathbb{Z}_m\times \mathbb{Z}_{2n}$-invariant $G^*(m,2n,4,3)$  design relative to $\{0\}\times \mathbb{Z}_{2n}$, which exists from the proof of Theorem \ref{S-ZmZnSQS(mn)-OOSPC}. Since there is a strictly semi-cyclic $G(2,2^{a-1}3^b,4,3)$  design by Lemma \ref{Semi-Cyclic-G(2,g,4,3)}, there is
strictly a $\mathbb{Z}_m\times \mathbb{Z}_{2^a3^bn}$-invariant $G(m,2^a3^bn,4,3)$  design relative to $\{0\}\times \mathbb{Z}_{2^a3^bn}$. Since there is an $S$-cyclic SQS$(2n)$ and an optimal $1$-D $(24,4,2)$-OOC with $J(1,24,4,2)-1$ codewords by Lemma \ref{Small-OOC} which is equivalent to a strictly cyclic PQS$(24)$, by Lemma \ref{(24n,4,2)-OOC} there is a strictly cyclic PQS$(2^a3^bn)$ with $J(1,2^a3^bn,4,2)-1$ base blocks.
Further, input this PQS into the strictly $\mathbb{Z}_{m}\times \mathbb{Z}_{2^a3^bn}$-invariant $G^*(m,2^a3^bn,4,3)$  design and apply Construction \ref{filling}.  We then obtain a strictly cyclic $\mathbb{Z}_{m}\times \mathbb{Z}_{2^a3^bn}$-invariant PQS$(2^a3^bmn)$ with $J(m,2^a3^bn,4,2)-1$ base blocks attaining the upper bound in Lemma $\ref{bound24k}$, which leads to an optimal $(m,2^a3^bn,4,2)$-OOSPC.  \qed

\begin{theorem}
\label{S-ZmZnSQS(mn)-OOSPC-2}
Let $m,n$ be equal to $1$ or odd integers such that there is an $S$-cyclic SQS$(2p)$ for each prime divisor $p$ of $m$ and $n$. Let $g\equiv 0 \pmod 8$. If there is an optimal $(2^{\epsilon},g,4,2)$-OOSPC with $J(2^{\varepsilon},g,4,2)$ codewords, then there is an optimal $(2^{\epsilon}m,ng,4,2)$-OOSPC with the size attaining the upper bound $(\ref{2D-JohnsonBound})$, where $\epsilon\in \{1,2\}$. If $2^{\epsilon}g\equiv 0\pmod {24}$ and there is an optimal $(2^{\epsilon},g,4,2)$-OOSPC with $J(2^{\varepsilon},g,4,2)-1$ codewords, then there is an optimal $(2^{\epsilon}m,ng,4,2)$-OOSPC with the size attaining the upper bound in Lemma {\rm \ref{bound24k}}, where $\epsilon\in \{1,2\}$.
\end{theorem}

\proof For $m=1$ and $n>1$, from the proof of Theorem \ref{S-ZmZnSQS(mn)-OOSPC}, there is a strictly $\mathbb{Z}_n\times \mathbb{Z}_{2^{\epsilon}}$-invariant $G(n,2^{\epsilon},4,3)$  design. Since there is a strictly semi-cyclic $G(2,g,4,3)$  design by Lemma \ref{Semi-Cyclic-G(2,g,4,3)}, Corollary \ref{G+semi-cyclic G(2,g,4,3)} shows that there is a strictly $\mathbb{Z}_{ng}\times \mathbb{Z}_{2^{\epsilon}}$-invariant $G(n,2^{\epsilon}g,4,3)$  design relative to $n\mathbb{Z}_{ng}\times \mathbb{Z}_{2^{\epsilon}}$. Since there is an $(2^{\epsilon},g,4,2)$-OOSPC with $J(2^{\epsilon},g,4,2)$ codewords by assumption, which is equivalent to a strictly $\mathbb{Z}_{g}\times \mathbb{Z}_{2^{\epsilon}}$-invariant PQS$(2^{\epsilon}g)$ with $J(2^{\epsilon},g,4,2)$ base blocks, applying Construction \ref{filling} gives a strictly $\mathbb{Z}_{ng}\times \mathbb{Z}_{2^{\epsilon}}$-invariant PQS$(2^{\epsilon}ng)$ with $J(ng,2^{\epsilon},4,2)$ base blocks. Therefore, there is an optimal $(2^{\epsilon},ng,4,2)$-OOSPC with the size attaining the upper bound $(\ref{2D-JohnsonBound})$ by Theorem \ref{equivalence}.

For $m>1$, from the proof of Theorem \ref{S-ZmZnSQS(mn)-OOSPC}, there is an $S$-cyclic SQS$(2^{\epsilon}m)$, which implies the existence of strictly cyclic $G(m,2^{\epsilon},4,3)$  design. Since there is a strictly semi-cyclic $G(2,ng,4,3)$  design by Lemma \ref{Semi-Cyclic-G(2,g,4,3)}, Corollary \ref{G+semi-cyclic G(2,g,4,3)} shows that there is a strictly $\mathbb{Z}_{2^{\epsilon}m}\times \mathbb{Z}_{ng}$-invariant $G(m,2^{\epsilon}ng,4,3)$  design relative to $m\mathbb{Z}_{2^{\epsilon}m}\times \mathbb{Z}_{ng}$.  Applying Construction \ref{filling} with the known strictly $\mathbb{Z}_{2^{\epsilon}}\times \mathbb{Z}_{ng}$-invariant PQS$(2^{\epsilon}ng)$ with $J(2^{\epsilon},ng,4,2)$ base blocks gives a strictly $\mathbb{Z}_{2^{\epsilon}m}\times \mathbb{Z}_{ng}$-invariant PQS$(2^{\epsilon}mng)$ with $J(2^{\epsilon}m,ng,4,2)$ base blocks. Therefore, there is an optimal $(2^{\epsilon}m,ng,4,2)$-OOSPC with the size attaining the upper bound $(\ref{2D-JohnsonBound})$.

When $2^{\epsilon}g\equiv 0\pmod {24}$, similar discussion as above gives the conclusion. \qed

\begin{theorem}
\label{(2m,8n,4,2)}
Let $m,n$ be equal to $1$ or the composite numbers of primes as in Corollary $\ref{(m,2n,4,2)}$ and $\epsilon\in \{1,2\}$. Then there is an optimal $(2^{\epsilon}m,8n,4,2)$-OOSPC with the size attaining the upper bound $(\ref{2D-JohnsonBound})$.
\end{theorem}

\proof For $m=n=1$, there is a $(2,8,4,2)$-OOSPC with $J(2,8,4,2)$ codewords \cite{Sawa2010}. By Example \ref{Z4Z8G(2,16,4,3)}, there is an optimal $(4,8,4,2)$-OOSPC with $J(4,8,4,2)$ codewords. For other values $m$ and $n$, applying Theorem \ref{S-ZmZnSQS(mn)-OOSPC-2} gives the conclusion.  \qed

Denote $U=\{4^r-1: r\ {\rm is\ a\ positive\ integer}\}\cup
\{1, 27, 33, 39, 51, 87, 123, 183\}$, and $P=\{p\equiv 7 \pmod {12}: p$ is a
prime$\}\cup \{2^n-1: {\rm odd \ integer} \ n\geq 1\}\cup \{25,
37, 61, 73, 109, 157, 181, 229, 277\}$, $V=\{v: v\in P$ or $v$ is a product of integers from the set $P\}$ and $M=\{uv: u\in U, v\in V\}\cup \{21^ru: r\geq 0, u\in \{3, 15, 21, 27, 33, 39, 51, 57, 63, 75, 87,
93\}\}$.

\begin{lemma}
\emph{\cite{FCJ2009}}
\label{known-RoSQS}
 There exists an $RoSQS(m+1)$ for $m\in M$.
\end{lemma}

\begin{theorem}
\label{(m,n,4,2)}
For $m,n\in M$, there is an optimal $(m,n,4,2)$-OOSPC with the size attaining the upper bound $(\ref{2D-JohnsonBound})$.
\end{theorem}

\proof If $m\equiv 1\pmod 6$, by Corollary \ref{RoSQS-OOSPC}, there is a strictly $\mathbb{Z}_m\times \mathbb{Z}_n$-invariant $1$-FG$(3,(3,4),mn)$ of type $n^{m}$ relative to $\{0\}\times \mathbb{Z}_n$. Construct a cyclic $1$-FG$(3,(3,4),n)$ on $\{0\}\times \mathbb{Z}_n$ which is obtained by deleting the fixed point. Then we obtain a $\mathbb{Z}_m\times \mathbb{Z}_n$-invariant $1$-FG$(3,(3,4),mn)$
type $1^{mn}$.  By Lemma \ref{1-FG-OOSPC}, there is  an optimal $(m,n,4,2)$-OOSPC.

If $m\equiv 3\pmod 6$, by Corollary \ref{RoSQS-OOSPC}, there is a strictly $\mathbb{Z}_m\times \mathbb{Z}_n$-invariant $1$-FG$(3,(3,4),mn)$ of type $(3n)^{m/3}$ relative to $\frac{m}{3}\mathbb{Z}_m\times \mathbb{Z}_n$. Also by Corollary \ref{RoSQS-OOSPC}, there is a strictly $\mathbb{Z}_n\times \mathbb{Z}_3$-invariant $1$-FG$(3,(3,4),mn)$ of type $3^{n}$ if $n\equiv 1\pmod 6$, and a strictly $\mathbb{Z}_n\times \mathbb{Z}_3$-invariant $1$-FG$(3,(3,4),3n)$ of type $9^{n/3}$ relative to $\frac{n}{3}\mathbb{Z}_n\times \mathbb{Z}_3$ if $n\equiv 3\pmod 6$. Since there is a $\mathbb{Z}_3\times \mathbb{Z}_3$-invariant $1$-FG$(3,(3,4),9)$ of type $3^{3}$ by Example \ref{1-FG(3,(3,4),9)} or from the proof of Theorem \ref{(p,p,p+1,2)-OOSPC}, there is a  $\mathbb{Z}_n\times \mathbb{Z}_3$-invariant $1$-FG$(3,(3,4),3n)$ of type $1^{3n}$ whenever $n\equiv 1\pmod 6$ or $n\equiv 3\pmod 6$. Therefore, there is
a $\mathbb{Z}_m\times \mathbb{Z}_n$-invariant $1$-FG$(3,(3,4),mn)$
type $1^{mn}$.  By Lemma \ref{1-FG-OOSPC}, there is  an optimal $(m,n,4,2)$-OOSPC. \qed

\begin{lemma}
\emph{\cite[Lemma 4.8]{FCJ2008}}
\label{Semi-Cyclic-G(3,g,4,3)}
There exists a strictly cyclic $G(3,g,4,3)$  design for each integer $g\equiv 0 \pmod {12}$.
\end{lemma}

\begin{lemma}
\emph{\cite[Corollary 6.14]{FCJ2008}}
\label{RoSQS-(v,4,2)-OOC}
If there is an RoSQS$(n+1)$, then for any integers $a,b\geq 0$, there is an optimal $1$-D $(3^a5^bn\cdot 36,4,2)$-OOC with the size attaining the upper bound $(\ref{2D-JohnsonBound})$.
\end{lemma}

\begin{theorem}
\label{(m,43{a+2}5bn,4,2)}
Let $m,n\in M$ and $a,b$ nonnegative integers. If $m\equiv 1\pmod 6$ then there is an optimal $(m,2^23^{a+2}5^bn,4,2)$-OOSPC with the size attaining the upper bound {\rm (\ref{2D-JohnsonBound})}.
\end{theorem}

\proof Since there is an RoSQS$(m+1)$ by Lemma \ref{known-RoSQS} and a strictly semi-cyclic $G(3,2^23^{a+2}5^bn,4,3)$  design by Lemma \ref{Semi-Cyclic-G(3,g,4,3)}, there is a strictly $\mathbb{Z}_m\times \mathbb{Z}_{2^23^{a+2}5^bn}$-invariant $G(m,2^23^{a+2}5^bn,4,3)$  design relative to $\{0\}\times \mathbb{Z}_{2^23^{a+2}5^bn}$ by Corollary \ref{RoSQS-ZmZn-G-desings}. By Lemma \ref{RoSQS-(v,4,2)-OOC}, there is an  optimal $1$-D $(3^a5^bn\cdot 36,4,2)$-OOC with $J(1,3^a5^bn\cdot 36,4,2)$ codewords, which is equivalent to a strictly PQS$(3^a5^bn\cdot 36)$. Applying Construction \ref{filling} gives an optimal $(m,3^a5^bn\cdot 36,4,2)$-OOSPC with the size attaining the upper bound (\ref{2D-JohnsonBound}). \qed

\begin{theorem}
\label{(m,4n,4,2)}
Let $m,n\in M$. If $m\equiv n\equiv 3\pmod 6$, then there is an optimal $(m,4n,4,2)$-OOSPC with the size attaining the upper bound in Lemma $\ref{Bound72k+18,36}$.
\end{theorem}

\proof Since there is an RoSQS$(m+1)$ by Lemma \ref{known-RoSQS} and a strictly semi-cyclic $G(3,4n,4,3)$  design by Lemma \ref{Semi-Cyclic-G(3,g,4,3)}, there is a strictly $\mathbb{Z}_{m}\times \mathbb{Z}_{4n}$-invariant $G(\frac{m}{3},12n,4,3)$  design relative to $\frac{m}{3}\mathbb{Z}_{m}\times \mathbb{Z}_{4n}$ by Corollary \ref{RoSQS-ZmZn-G-desings}. Since there is an RoSQS$(n+1)$ by assumption  and a strictly semi-cyclic $G(3,12,4,3)$  design by Lemma \ref{Semi-Cyclic-G(3,g,4,3)}, there is a strictly $\mathbb{Z}_{n}\times \mathbb{Z}_{12}$-invariant $G(n/3,36,4,3)$  design relative to $\frac{n}{3}\mathbb{Z}_{n}\times \mathbb{Z}_{12}$ by Corollary \ref{RoSQS-ZmZn-G-desings}, thereby, there is a strictly $\mathbb{Z}_{4n}\times \mathbb{Z}_{3}$-invariant $G(n/3,36,4,3)$  design $\frac{n}{3}\mathbb{Z}_{4n}\times \mathbb{Z}_{3}$. Since there is an optimal $(3,12,4,2)$-OOSPC with the size attaining the upper bound in Lemma $\ref{Bound72k+18,36}$, there is a strictly $\mathbb{Z}_{4n}\times \mathbb{Z}_{3}$-invariant PQS$(12n)$ with $J(4n,3,4,2)-1$ base blocks. Applying Construction \ref{filling} gives optimal $(m,4n,4,2)$-OOSPC with the size attaining the upper bound in Lemma $\ref{Bound72k+18,36}$. \qed

\begin{theorem}
\label{(m,2{a+2}3{b+1}n,4,2)}
Let $m\in M$ and $n$ be composite numbers of primes as in Corollary $\ref{(m,2n,4,2)}$ and  let $a,b$ be two nonnegative integers.  If $m\equiv 1\pmod 6$, then there is an optimal $(m,2^{a+2}3^{b+1}n,4,2)$-OOSPC with the size attaining the upper bound in Lemma $\ref{bound24k}$.
\end{theorem}

\proof  Since there is an RoSQS$(m+1)$ by Lemma \ref{known-RoSQS} and a strictly semi-cyclic $G(3,2^{a+2}3^{b+1}n,4,3)$  design by Lemma \ref{Semi-Cyclic-G(3,g,4,3)}, there is a strictly $\mathbb{Z}_m\times \mathbb{Z}_{2^{a+2}3^{b+1}n}$-invariant $G(m,2^{a+2}3^{b+1}n,4,3)$  design relative to  $\{0\}\times \mathbb{Z}_{2^{a+2}3^{b+1}n}$ by Corollary \ref{RoSQS-ZmZn-G-desings}. Since $n$ is a composite number of primes as in Corollary $\ref{(m,2n,4,2)}$, there is a cyclic SQS$(2n)$ by Theorem \ref{S-CSQS}.  By Lemma \ref{(24n,4,2)-OOC}, there is an  optimal $1$-D $(2^{a+2}3^{b+1}n,4,2)$-OOC with the size attaining the upper bound in Lemma \ref{bound24k}, which is equivalent to a strictly cyclic PQS$(2^{a+2}3^{b+1}n)$. Applying Construction \ref{filling} gives an optimal $(m,2^{a+2}3^{b+1}n,4,2)$-OOSPC with the size attaining the upper bound in Lemma \ref{bound24k}.
\qed

\section{Concluding Remark}

In this paper, we gave some combinatorial constructions for optimal $(m,n,4,2)$-OOSPCs. As applications, many infinite families of optimal $(m,n,4,2)$-OOSPCs were obtained. We summarized all infinite families obtained in this table I. As pointed out in the remark of Lemma \ref{(m,2n,4,2)}, Sawa's result in \cite{Sawa2010} was obtained again and our construction seemed easier. We also obtained some infinite classes of optimal $(m,n,4,2)$-OOSPCs with $gcd(m,n)$ being divisible by 2 or 3. Our constructions strength the importance of $S$-cyclic SQSs and RoSQSs. They are worth studying.

\begin{figure*}[!t]
% ensure that we have normalsize text
\normalsize
% Store the current equation number.
%\setcounter{mytempeqncnt}{\value{equation}}
% Set the equation number to one less than the one
% desired for the first equation here.
% The value here will have to changed if equations
% are added or removed prior to the place these
% equations are referenced in the main text.

\centerline{\footnotesize Table I}
\centerline{\footnotesize New Infinite families of optimal $(m,n,k,2)$-OOSPCs}
\vspace{0.1cm}
\begin{center}
\begin{tabular}{|c|c|c|c|}
\hline
Parameters & Conditions & Size & Source \\ \hline
$(p,p,p+1,2)$ & $p$ is a prime & $J(p,p,p+1,2)$ & Theorem \ref{(p,p,p+1,2)-OOSPC}\\ \hline
$(m,2ng,4,2)$ & $m, n\in W$ & $J(m,2ng,4,2)$ & Corollary \ref{(m,2n,4,2)}\\
$(mg,2n,4,2)$ & $g\in \{1,3,5,\ldots, 49\}\setminus \{33\}$ & $J(mg,2n,4,2)$ & \\ \hline
$(m,4ng,4,2)$ & $m, n\in W$ & $J(m,4ng,4,2)$ & Corollary \ref{(m,2n,4,2)}\\
$(mg,4n,4,2)$ & $g\in \{1,3,5,\ldots, 13\}\setminus \{3\}$ & $J(mg,4n,4,2)$ & \\ \hline
$(3m,bn,4,2)$ & $m,n\in W$, $b\in \{6,12\}$ & $J(3m,bn,4,2)-1$ & Theorem \ref{(3m,bn,4,2)}\\ \hline
$(m,2^a3^bn,4,2)$ & $m,n\in W$, $a\geq 4,b>1$ & $J(m,2^a3^bn,4,2)-1$ & Theorem \ref{(m,2a3bn,4,2)} \\ \hline
$(2^{\epsilon}m,8n,4,2)$ & $m,n\in W$, $\epsilon\in \{1,2\}$ & $J(2^{\epsilon}m,8n,4,2)$ & Theorem \ref{(2m,8n,4,2)}\\ \hline
${(m,n,4,2)}$ & $m,n\in M$ & $J{(m,n,4,2)}$ & Theorem \ref{(m,n,4,2)}\\ \hline
$(m,2^23^{a+2}5^bn,4,2)$ & $m,n\in M$, $m\equiv 1\pmod 6$, & $J(m,2^23^{a+2}5^bn,4,2)$ & Theorem \ref{(m,43{a+2}5bn,4,2)}\\ & $a,b\geq 0$ & & \\ \hline
${(m,4n,4,2)}$ & $m,n\in M$, $m,n\equiv 3\pmod 6$ & $J{(m,4n,4,2)}-1$ & Theorem \ref{(m,4n,4,2)}\\ \hline
$(m,2^{a+2}3^{b+1}n,4,2)$ & $m\in M$, $m\equiv 1\pmod 6$, & $J(m,2^{a+2}3^{b+1}n,4,2)-1$ & Theorem \ref{(m,2{a+2}3{b+1}n,4,2)} \\ &  $n\in W$, $a,b\geq 0$ & & \\ \hline
\end{tabular}
\end{center}

{\small
$W=\{p_1^{a_1}p_2^{a_2}\cdots p_r^{a_r}: {\rm each\ prime}\ p_i\equiv 1\pmod{12}\ {\rm and}\ p_i<10^5,\ {\rm or}\ p_i\equiv 5\pmod{12}\ {\rm and}\ p<1500000\};$

\smallskip
$M=\{uv: u\in U, v\in V\}\cup \{21^ru: r\geq 0, u\in \{3, 15, 21, 27, 33, 39, 51, 57, 63, 75, 87,
93\}\}$, where $U=\{4^r-1: r\ {\rm is\ a\ positive\ integer}\}\cup
\{1, 27, 33, 39, 51, 87, 123, 183\}$, and $P=\{p\equiv 7 \pmod {12}: p$ is a
prime$\}\cup \{2^n-1: {\rm odd \ integer} \ n\geq 1\}\cup \{25,
37, 61, 73, 109, 157, 181, 229, 277\}$, $V=\{v: v\in P$ or $v$ is a product of integers from the set $P\}$.
}
% Restore the current equation number.
%\setcounter{equation}{\value{mytempeqncnt}}
% IEEE uses as a separator
%\hrulefill
% The spacer can be tweaked to stop underfull vboxes.
\vspace*{4pt}
\end{figure*}

By Lemma \ref{bound24k} and Lemma \ref{Bound72k+18,36}, we see that the Johnson bound can not be achieved in some cases. The problem of constructing optimal $(m, n, w, \lambda)$-OOSPC is apparently a
difficult and challenging task in general weight.
Although the case of $w = 4$ is too small
for practical application, we hope that it may help us to study the other larger cases.

%\medskip
%
%\section*{Acknowledgment}
%
%The authors are very grateful to the reviewers for their valuable
%comments and suggestions that improved the presentation.

\medskip

{\bf \centerline {APPENDIX A}}

{\em Proof of Construction \ref{G*-design}}:
Firstly, we compute the number of blocks in ${\cal C}'$. Since $m-e$ is even, the cardinality of $\mathbf{I}$ is $\frac{m-e}{2}$.
It is easy to
see that $|\C_1|=g^2|\F_2|=g^2 n(m-e)(mn+en-9)/24$,
$|\C_2|=ng(m-e)/4$, $|\C_3|=ng(g-1)(m-e)/4$ and
$|\C_4|=ng(m-e)(g-1)/8$. Thus,
$$|{\cal C}'|=mng(|\C_1|+|\C_2|+|\C_3|+|\C_4|)=\frac{n^2g^2m(m-e)(mng+eng-3)}{24},$$
which is the expected number of
quadruples. Also, it is $\mathbb{Z}_{m}\times \mathbb{Z}_n$-invariant, thereby, it
suffices to show that each triple containing $(0,0)$ and not contained in any group appears
in at least one
 quadruple of ${\cal C}'$.
 Let $T=\{(0,0),(x_1,y_1+z_1n),(x_2,y_2+z_2n)\}$ be such a triple, where $x_k\in \mathbb{Z}_{m}$, $0\leq y_k<n$, $0\leq z_k<g$ and $1\leq k\leq 2$. Clearly, at most one of $(x_1,y_1)$ and $(x_2,y_2)$ belongs to $(\frac{m}{e}\mathbb{Z}_{m})\times \mathbb{Z}_n$. The proof is divided into two cases.

Case 1: Two of $(0,0),(x_1,y_1)$ and $(x_2,y_2)$ are equal. When $(x_1,y_1)=(x_2,y_2)$, we have $x_1\not\in \frac{m}{e}\mathbb{Z}_m$. If $x_1\in \mathbf{I}$ then there is a block $B=\{(0,0),(x_1,y_1+z_1n),(x_1,y_1+z_2n),(2x_1,2y_1+z_1n+z_2n)\}\in \C_3\subset \C_3'$ such that $T\subset B$.
If $x_1\not\in \mathbf{I}$, then $-x_1\in \mathbf{I}$, thereby there is a base block $B=\{(0,0),(-x_1,-y_1-z_1n),(-x_1,-y_1-z_2n),(-2x_1,-2y_1-z_1n-z_2n)\}\in \C_3$ such that $T\subset B+(2x_1,2y_1+z_1n+z_2n)\in \C_3'$. When one of $(x_1,y_1)$ and $(x_2,y_2)$ is equal to $(0,0)$, without loss of generality let $(x_1,y_1)=(0,0)$, consider the triple $T-(x_2,y_2+z_2n)$. Similarly, there is a block $C\in \C'$ such that $T-(x_2,y_2+z_2n)\subset C$, thereby, there is a block $C'\in \C'$ containing $T$.

Case 2: $(0,0),(x_1,y_1)$ and $(x_2,y_2)$ are distinct. Since $(\mathbb{Z}_{m}\times \mathbb{Z}_n,\{\{i,i+\frac{m}{e},\ldots,i+m-\frac{m}{e}\}\times \mathbb{Z}_n:0\leq i<\frac{m}{e}\},{\cal B})$ is a $\mathbb{Z}_{m}\times \mathbb{Z}_n$-invariant $G^*(\frac{m}{e},en,4,3)$  design, there is a base block $B=\{(x_1',y_1'),(x_2',y_2'),(x_3',y_3')$, $(x_4',y_4')\}\in \F$ and $(\tau,\mu)\in \mathbb{Z}_{m}\times \mathbb{Z}_n$ such that $\{(0,0),(x_1,y_1), (x_2,y_2)\}\subset B+(\tau,\mu)$. Without loss of generality, let $(x_k,y_k)=(x_k',y_k')+(\tau,\mu)$ for $1\leq k\le 2$ and $(0,0)=(x_3',y_3')+(\tau,\mu)$. If $B\in \F_2$ then let $y_k'+\mu=a_kn+y_k$, $a_k\in \{0,1\}$ for $1\leq k\leq 3$ where $y_3=0$. Since  $\A_B$ is the set of base blocks of a semi-cyclic $H(4,g,4,3)$  design on $\{(x_l',y_k'+un): 1\leq k\leq 4, 0\leq u<g\}$ with groups $\{(x_k',y_k'+un):0\leq u<g\}$ ($1\leq k\leq 4$), there is a unique base block $A=\{(x_1',y_1'+z_1'n),(x_2',y_2'+z_2'n),(x_3',y_3'+z_3'n),(x_4',y_4'+z_4'n)\}\in \A_B$ and an element $\rho\in \{0,1,\ldots,g-1\}$ such that $\{(x_1',y_1'+z_1n-a_1n),(x_2',y_2'+z_2n-a_2n)$,$(x_3',y_3'+z_3n-a_3n)\}\subset A+(0,\rho n)$. It follows that $T\subset A+(\tau,\mu+\rho n)\in {\cal C}_1'.$
If $B\in \F_1$, then $\{(0,0),(x_1,y_1), (x_2,y_2)\}$ is of the form $\{(-x,-y),(0,0),(x,y)\}+(\tau,\mu)$ where $(x,y)\in \mathbf{I}\times \mathbb{Z}_n$, or of the form $\{(0,0),(x,y),(0,\frac{n}{2})\}+(\tau,\mu)$ where $(x,y)\in (\mathbb{Z}_m\setminus \frac{m}{e}\mathbb{Z}_m)\times \{0,1,\ldots,\frac{n-2}{2}\}$.

Suppose that $\{(0,0),(x_1,y_1), (x_2,y_2)\}$ is of the form $(\tau,\mu)+\{(0,0),(x,y),(-x,-y)\}$, $x\in \mathbf{I}, 0\leq y<n$. Then there is a triple of the form $\{(0,0),(x,y+zn),(-x,-y+z'n)\}$ ($0\leq z,z'<g$) in the orbit generated by $\{(0,0),(x_1,y_1+z_1n),(x_2,y_2+z_2n)\}$ under $\mathbb{Z}_{m}\times \mathbb{Z}_n$. If $z'=-z$ then $\{(0,0),(x,y+zn),(-x,-y+z'n)\}\subset \{(0,0),(x,y+zn),(-x,-y+z'n),(0,\frac{gn}{2})\}\in \C_2$. Otherwise,  $\{(0,0),(x,y+zn),(-x,-y+z'n)\}\subset \{(0,0),(x,y+zn),(x,y+(g-z')n),(2x,2y+zn-z'n)\}-(x,y-z'n)\in \C_3'$. It follows that $T$ occurs in a block of $\C_2'\cup \C_3'$.

Suppose that $\{(0,0),(x_1,y_1), (x_2,y_2)\}$ is of the form $(\tau,\mu)+\{(0,0),(x,y),(0,\frac{n}{2})\}$, $x\in \mathbb{Z}_{m}\setminus \frac{m}{e}\mathbb{Z}_{m}, 0\leq y<\frac{n}{2}$. Then there is a triple of the form $\{(0,0),(x,y+zn),(0,\ell n+\frac{n}{2})\}$ ($0\leq z,\ell<g$) in the orbit generated by $\{(0,0),(x_1,y_1+z_1n),(x_2,y_2+z_2n)\}$ under $\mathbb{Z}_{m}\times \mathbb{Z}_n$.  For $x\in \mathbf{I}$, if $0\leq \ell<\frac{g-1}{2}$ then $\{(0,0),(x,y+zn),(0,\ell n+\frac{n}{2})\}\subset \{(0,0),(0,\ell n+\frac{n}{2}),(x,y+zn),(x,y+\frac{n}{2}+\ell n+z n)\}\in \C_4$, if $\ell=\frac{g-1}{2}$ then
$\{(0,0),(x,y+zn),(0,\ell n+\frac{n}{2})\}\subset \{(0,0),(x,y+zn),(-x,-y-zn),(0,\ell n+\frac{n}{2})\}\in \C_2$ or $\{(0,0),(x,y+zn),(0,\ell n+\frac{n}{2})\}\subset \{(0,0),(x,y+zn),(-x,-y-zn),(0,\ell n+\frac{n}{2})\}+(0,\frac{ng}{2})\in \C_2'$ according to $0\leq y+zn<\frac{ng}{2}$ or not,
otherwise, $\{(0,0),(x,y+zn),(0,\ell n+\frac{n}{2})\}\subset \{(0,0),(0,(g-\ell-1)n+\frac{n}{2}),(x,y-\frac{n}{2}+zn-\ell n), (x,y+zn)\}+(0,\frac{n}{2}+ln)\in \C_4'$.
It follows that $T$ occurs in a block of $\C_4'$. For $-x\in \mathbf{I}$, similar discussion shows that $T$ is contained a block of ${\cal C}'$. \qed

\medskip

{\bf \centerline {APPENDIX B}}

{\em Proof of Construction \ref{FC-sFG}}:
For checking the required design, it suffices to show that: ($1$)
the resulting design is strictly $\mathbb{Z}_{m}\times \mathbb{Z}_{ng}$-invariant; ($2$) any triple $T$,
$T\subset \mathbb{Z}_m\times \mathbb{Z}_{ng}$, $|T \cap G'|<3$ for all $G'\in {\cal G}'$, is
contained in a unique block of the resulting design; ($3$) any
pair of points $P$, $P\subset \mathbb{Z}_m\times \mathbb{Z}_{ng}$, $|P \cap G'|<2$ for all $G'\in {\cal
G}'$, is contained in a unique block of ${{\cal A}'_j}$ for each
$0\leq j< s$.

($1$) Suppose that $A=\{(x_1,y_1+z_1n),(x_2,y_2+z_2n),\ldots,(x_r,y_r+z_rn)\}$ is a
base block of the resulting design, where $x_l\in \mathbb{Z}_m$,
$0\leq y_l\leq n-1$, $0\leq z_l\leq g-1$, $1\leq l\leq r$. We need to show that the
stabilizer of $A$ is trivial, i.e. $A+\delta= A$ if and only if
$\delta=(0,0)$. The sufficiency follows immediately,
so we consider the necessity.

 Assume that $\delta=(\delta_1,\delta_2+\delta_3n)$, $\delta_1\in \mathbb{Z}_m$, $0\leq \delta_2\leq n$, $0\leq \delta_3<g$. If
 $A+\delta=A$ then
$$\{(x_l,y_l+z_ln): 1\leq l\leq r\} =
\{(x_l+\delta_1,y_l+z_ln+\delta_2+\delta_3n): 1\leq l\leq r\},$$ where the
arithmetic is in the ring $\mathbb{Z}_m\times \mathbb{Z}_{ng}$. It follows that
$$\{(x_l,y_l): 1\leq l\leq r\} = \{(x_l+\delta_1,y_l+\delta_2): 1\leq l\leq
r\},$$ where the arithmetic is in the ring $\mathbb{Z}_m\times \mathbb{Z}_n$. Let $U=\{(x_l,y_l): 1\leq
l\leq r\}$.

If $A\in {{\cal A}'_j}$, $0\leq j< s$, then $|U|=r\geq 2$. Since
the subdesign $(X,{\cal G},{\cal B}_0)$ of the master design
$1$-FG$(3,(K_0,K_1),mn)$ of type ${(en)}^{m/e}$ $(\mathbb{Z}_m\times \mathbb{Z}_n,{\cal G},{\cal B}_0,{\cal
B}_1)$ is strictly $\mathbb{Z}_m\times \mathbb{Z}_n$-invariant and it requires that any $2$-subset of $\mathbb{Z}_m\times \mathbb{Z}_n$
which intersects any group of ${\cal G}$ in at most one point occurs
in exactly one block, we have $(\delta_1,\delta_2)=(0,0)$.

If $A\in {\cal A}_s'$, without loss of generality we can always assume
that $A\in {\cal A}_s^*$. If $A=\tau(C)$ for some $C\in
\bigcup_{B\in{\cal F}_1} {{\cal D}_B}$, then $|U|=r\geq 3$. Since
the master design $1$-FG$(3,(K_0,K_1),mn)$ of type $(en)^{m/e}$ is strictly
$\mathbb{Z}_m\times \mathbb{Z}_n$-invariant and it requires that any $3$-subset of $\mathbb{Z}_m\times \mathbb{Z}_n$ which intersects
any group of ${\cal G}$ in at most two points occurs in exactly one
block, we have $(\delta_1,\delta_2)=(0,0)$. If $A=\tau(C)$ for some $C\in
\bigcup_{B\in{\cal F}_0} {{\cal A}_B^s}$, then $|U|\geq 2$. Note that
in this case $U$ may be a multiset, i.e. $|U|$ may be not equal to
$r$. By similar arguments as the case of $A\in {{\cal A}'_j}$, we
have $(\delta_1,\delta_2)=(0,0)$.

Hence,
$$\{(x_l,y_l+z_ln): 1\leq l\leq r\} =
\{(x_l,y_l+z_ln+\delta_3n): 1\leq l\leq r\},$$
where the arithmetic is
in the ring $\mathbb{Z}_m\times \mathbb{Z}_{ng}$. Since the input designs are all strictly semi-cyclic,
we have $\delta_3=0$. Thus the resulting design is strictly $\mathbb{Z}_m\times \mathbb{Z}_{ng}$-invariant.

($2$) Take any triple $T=\{(x_1,y_1+z_1n),(x_2,y_2+z_2n),(x_3,y_3+z_3n)\}\subset \mathbb{Z}_m\times \mathbb{Z}_{ng}$ which is not contained in any group of $\G'$, where
$x_l\in \mathbb{Z}_m$,
$0\leq y_l\leq n-1$, $0\leq z_l\leq g-1$, $1\leq l\leq 3$ and $x_1,x_2,x_3$ are not congruent to the same number modulo $\frac{m}{e}$.  We consider the following cases.

Case $1$. Suppose that $x_1$, $x_2$, $x_3$ are pairwise distinct
modulo $\frac{m}{e}$. Then there exists a unique base block $B$ in ${\cal F}$
and unique elements $\delta_1,\delta_2$ with $0\leq \delta_1<m$ and $0\leq \delta_2<n$, such that
$\{(x_1,y_1),(x_2,y_2),(x_3,y_3)\}\subseteq B+(\delta_1,\delta_2)$. Let $(x_l^*,y_l^*)\in B$ satisfy $x_l\equiv x_l^*+\delta_1\pmod m$ and $y_l^*+\delta_2=y_l+\sigma_l
n$ for some $\sigma_l\in \{0,1\}$, $1\leq l\leq 3$. Note that $x_1^*$, $x_2^*$,
$x_3^*$ are also pairwise distinct modulo $\frac{m}{e}$.

If $B\in {\cal F}_0$, then there exists a unique base block $C\in
{\cal A}_B$ and a unique element $\delta_3$ with $0\leq \delta_3<g$, such that
$\{(x_1^*,y_1^*,z_1^*),(x_2^*,y_2^*,z_2^*),(x_3^*,y_3^*,z_3^*)\}\subseteq C$ and
$(x_l^*,y_l^*, z_l^*+\delta_3)=(x_l^*,y_l^*,z_l-\sigma_l+{\sigma'_l}g)$ for some
${\sigma'_l}\in \{0,1\}$, $1\leq l\leq 3$. By the mapping $\tau$, we
have that
$(x_l^*,y_l^*+z_l^*n)+(\delta_1,\delta_2+\delta_3n)=(x_l^*+\delta_1,y_l^*+\delta_2+z_l^*n+\delta_3n)=(x_l,y_l+\sigma_ln+z_l^*n+\delta_3n)=(x_l,y_l+z_ln)$.
Let $\delta=(\delta_1,\delta_2+\delta_3n)$. By
$(1)$ the resulting design is strictly $\mathbb{Z}_m\times \mathbb{Z}_{ng}$-invariant, so $T$
is contained in the unique block $\tau(C)+\delta$, which is
generated by $\tau(C)$. Similar arguments show that if  $B\in {\cal
F}_1$ then there is a unique base block $C\in \D_B$ and a unique element $\delta\in \mathbb{Z}_m\times \mathbb{Z}_{ng}$ such that $T\subset \tau(C)+\delta$.

Case $2$. Suppose that $x_1\equiv x_2\not \equiv x_3\pmod {\frac{m}{e}}$, $x_1\neq
x_2$. By similar
arguments as in Case $1$, there exists a unique base block $B\in
{\cal F}_1$, a unique base block $C\in {{\cal D}_B}$ and unique elements $\delta_1,\delta_2,\delta_3$ with $0\leq \delta_1<m$, $0\leq \delta_2<n$ and $0\leq \delta_3<g$, such that $T$ is contained in the unique block
$\tau(C)+\delta$, where $\delta=(\delta_1,\delta_2+\delta_3n)$, which is
generated by $\tau(C)$.

Case $3$. Suppose that $x_1=x_2$, $y_1\neq y_2$ and $x_1\not \equiv
x_3\pmod {\frac{m}{e}}$. By similar arguments as in Case $1$, there
exists a unique base block $B\in{\cal F}_0$, a unique base block
$C\in {\cal A}_B$ and unique elements $\delta_1,\delta_2,\delta_3$ with $0\leq \delta_1<m$, $0\leq \delta_2<n$ and $0\leq \delta_3<g$, such that $T$ is contained in the unique block
$\tau(C)+\delta$, where $\delta=(\delta_1,\delta_2+\delta_3n)$, which is
generated by $\tau(C)$.

($3$) Take any $2$-subset $P=\{(x_1,y_1+z_1n),(x_2,y_2+z_2n)\}$ which is not contained in any group of $\G'$, where
$x_l\in \mathbb{Z}_m$,
$0\leq y_l\leq n-1$, $0\leq z_l\leq g-1$, $1\leq l\leq 2$.  Then $x_1\not\equiv x_2\pmod {\frac{m}{e}}$ and there exists a unique
base block $B$ in ${\cal F}_0$ and unique elements $\delta_1,\delta_2$ with $0\leq \delta_1<m$ and $0\leq \delta_2<n$, such that $\{(x_1^*,y_1^*),(x_2^*,y_2^*)\}\subseteq B$ and
$x_l^*+\delta_1\equiv x_l\pmod m$ and $y_l^*+\delta_2=y_1+\sigma_ln$ for some $\sigma_l\in \{0,1\}$,
$1\leq l\leq 2$. Note that $x_1^*$, $x_2^*$
are also distinct modulo $\frac{m}{e}$.

Then, given any $0\leq j< s$, there exists a unique base block
$C_j$ in ${\cal A}_j^*$ and a unique element $\delta_{3}\in \mathbb{Z}_{g}$,
such that $\{(x_1^*,y_{1}^*,z_{1}^*),(x_2^*,y_2^*,z_{2}^*)\}\subseteq C_j$ and
$(x_l^*,y_l^*,z_{l}^*+\delta_{3})=(x_l^*,y_l^*,z_l-\sigma_l+{\sigma'_{l}}g)$
for some ${\sigma'_{l}}\in \{0,1\}$, $1\leq l\leq 2$. By the
mapping $\tau$, we have that
$(x_l^*+\delta_1,y_l^*+\delta_2+z_{l}^*n+\delta_{3}n)=(x_l,y_l+z_ln)
$. Let $\delta=(\delta_1,\delta_2+\delta_{3}n)$. By
$(1)$ the resulting design is strictly cyclic, so $P$ is
contained in the unique block $\tau(C_j)+\delta$, which is generated
by $\tau(C_j)$.

So, $(\mathbb{Z}_{m}\times \mathbb{Z}_{ng},{\cal G}',{{\cal A}'_0},\ldots,{{\cal
A}'_s})$ is a strictly $\mathbb{Z}_{m}\times \mathbb{Z}_{ng}$-invariant
$s$-FG$(3,(L_0,\ldots,L_s),mng)$ of type $(eng)^{m/e}$. \qed

\end{document}